\documentclass[]{JFM-FLM-arxiv}

\usepackage{bm}
\usepackage{soul}

\newcommand\Req{R_{\mathrm{eq}}}
\newcommand\Rmax{R_{\mathrm{max}}}

\definecolor{x}{rgb}{0.01, 0.75, 0.24}

\newcommand{\ssf}{\mathsfbi{S}}
\newcommand{\osf}{\boldsymbol{\mathsf{\Omega}}}

\newcommand{\pst}{P_{\mathrm{st}}}
\newcommand{\G}{\mathsfbi{G}}
\newcommand{\W}{\mathsfbi{W}}
\newcommand{\Rbr}{R_{\mathrm{br}}}
\newcommand{\Tbr}{T_{\mathrm{br}}}

\newcommand{\brac}[1]{\left( #1 \right) }

\lefttitle{ }
\righttitle{ }

\title{Stretching and breaking of particles in \\compressible random flows}

\author{Dipankar Roy\aff{1}, Marco Martins Afonso\aff{2}, Jason R.~Picardo\aff{3}, \and Dario Vincenzi\aff{1}}

\affiliation{\aff{1}Universit\'e C\^ote d'Azur, CNRS, LJAD, Nice 06100, France
\aff{2}SIT Technologies, Via Montallegro 1, 16145 Genoa, Italy
\aff{3}Department of Chemical Engineering, Indian Institute of Technology Bombay, Mumbai, 400076, India}

\corresau{Jason R. Picardo, \email{picardo@iitb.ac.in}}

\begin{document}
\maketitle

\begin{abstract}
A key feature of turbulent suspensions that involve floating particles on the surface or inertial particles in the bulk is the compressibility of the effective particle-phase velocity field.
Little, however, is known about the effects of small-scale flow compressibility on the stretching and breaking of particles. Here, we gain insight into the nature of these effects by studying the deformation of
tiny particles in model fluctuating flows. We consider a generic 
particle with extensional dynamics that are governed by a vector model, which accounts for elasticity, internal viscosity, and non-affine deformation. Applying the dynamical systems approach of \citet*{bfl00}, we first obtain general results for the stationary statistics of particle extension in compressible chaotic flows. We then specialize to a time-decorrelated Gaussian random flow and derive an exact solution for 
the Batchelor regime of the compressible Kraichnan model. We also perform numerical simulations for a time-correlated renewing flow. While straining is suppressed on the average in compressible flows, our results show that large deviations of the strain rate strongly stretch particles and give rise to a power-law distribution of extensions. Extreme straining events are particularly important for stiff particles and, in the examples considered here, give rise to a counter-intuitive effect: stiff particles stretch more and break faster in flows of increasing compressibility. Highly-elastic particles, whose deformation is dictated by the mean straining, stretch less and break slower. As a consequence, the shrink--stretch transition that occurs on varying the particle's relaxation time becomes increasingly shallow with compressibility, ultimately vanishing in the strongly compressible regime (wherein line elements contract on the average). 
Though based on specific random flows, our work shows how compressibility can affect the extensional dynamics of particles by altering the fluctuations of the strain rate, including its large deviations. 
\end{abstract}



\section{Introduction}
\label{sect:intro}

The dynamics of stretchable particles suspended in a fluid, such as droplets, macromolecules, or elastic fibers, depends strongly on the nature of the flow that transports them.
Indeed, several classification schemes for laminar flows are based on the flow's ability to distort a test particle (\citealt{tanner76}; \citealt*{orl82}; \citealt{2023-cunha-etal}).
For turbulent flows, the study of particle stretching relies on a statistical approach.
In particular, the deformation of a tiny stretchable particle is closely related to that of a material line element and, thus, may be analyzed using techniques from the theory of Lagrangian chaos (\citealt{cfvp91}; \citealt{bjpv98}; \citealt*{ccv2010}).
This approach has been used, in the context of incompressible turbulent flows, to predict the statistics of stretching and breakup of polymers (\citealt*{bfl00,bfl01}; \citealt{c00,vwrp21}) and droplets \citep*{bmv14,2018-ray-vincenzi}. 
Our goal is to extend these results to compressible random flows, while considering a generic stretchable particle.

One scenario of interest is that of deformable particles confined to the free surface of an incompressible fluid in turbulent motion. The chaotic two-dimensional flow at the surface has compressible small-scale dynamics---the contraction of area elements at the surface produces strong inhomogeneities in the spatial distribution of floating particles (\citealt{so93,cdgs04}; \citealt*{bgh04}; \citealt{coletti23}). In three dimensions, our study would be relevant to tiny dense particles whose motion departs weakly from that of fluid tracers; such particles can be assigned an effective velocity field that is compressible \citep{Boffetta07-eulerian,Haller08}, with a negative (positive) divergence in regions where the underlying fluid flow is strain-dominated (vortical)~\citep*{ravichandran-rev2017,bec21-dusty}. A situation that is \textit{not} of interest here is high-Mach-number turbulence. In this case, indeed, the small-scale dynamics are effectively incompressible \citep{fouxon2023}: while the instantaneous velocity gradient has a non-zero trace, the three Lyapunov exponents sum to zero, implying that volume elements 
maintain their volume on the average. 


To appreciate how compressibility at small scales affects the deformation of particles, it is helpful to first recall the key concepts of the general theory of line-element stretching in chaotic flows. 
The deformation of a material line element $\bm\ell(t)$, that is transported in a flow with velocity field $\bm u$, is determined by the velocity gradient $\bnabla{\bm u}$ sampled along a Lagrangian trajectory $\bm x(t)$: $\mathrm{d}_t \ell_i = \ell_j\,\partial_j u_i(\bm x(t),t)$. Let us denote the instantaneous rate of exponential stretching or contraction as $\gamma(t)$, so that $\ell(t)/\ell(0) = \exp(\gamma t)$ where $\ell(t)=\vert\bm\ell(t)\vert$. 
The long-time average of $\gamma(t)$ is the principal Lyapunov exponent $\lambda = \lim_{t\to\infty} \langle \gamma(t)\rangle$,
where $\langle\boldsymbol{\cdot}\rangle$ denotes the average over the statistical realizations of the velocity field.
Information on the fluctuations of $\gamma(t)$, and hence on the statistical deviations from the average dynamics, is provided by the \textit{generalized} Lyapunov exponents $L(q)$, which determine the long-time evolution of moments of $\ell(t)$:
$\langle \ell^q(t)\rangle \sim \ell^q(0)\mathrm{e}^{L(q) t}$ as $t\to\infty$ (\citealt{zrms84,bppv85}; see also \citealt{cfvp91,bjpv98,ccv2010}). 
The function $L(q)$ is
convex and satisfies $L(0)=0$ and $L'(0)=\lambda$. Moreover, if $\lambda>0$, then $L(-d_c)=0$ with $d_c$ being the correlation dimension of the attractor of fluid-particle trajectories \citep{b91}. The generalized Lyapunov exponents are related, via a Legendre transform, to the rate function (also known as the  Cram\'er function), which determines the form of the large-deviations approximation to the probability density function of $\gamma(t)$ \citep{bjpv98}.
In general, $L(q)$ is difficult to calculate analytically and is known explicitly only for a few random flows, including incompressible linear renewing flows \citep{y99}, telegraph noise \citep{fma07}, and isotropic Gaussian flows with vanishing correlation time \citep*{ckv98}.
A quartic polynomial approximation of $L(q)$ valid for general random flows is derived in \citet*{fak19}.

In incompressible flows, $\lambda$ is positive and all positive-order moments of $\ell(t)$ grow exponentially in time. Further, the correlation dimension
 $d_c$ equals the spatial dimension of the flow, i.e. material points or tracers sample the fluid uniformly. $L(q)$ or, equivalently, the rate function has been computed in incompressible turbulent flows, specifically in numerical simulations of isotropic turbulence \citep{bbbcmt06,jm15,jm16} and turbulent channel flows \citep{bmpb12,jhbm17}. 

Now, let us consider compressible flows. A measure of the Eulerian degree of compressibility of a velocity field $\bm u$ is provided by the ratio
\begin{equation}
\wp=
\dfrac{\langle(\bnabla\boldsymbol{\cdot}\bm u)^2\rangle}{\langle\Vert\bnabla\bm u\Vert^2\rangle},
\label{eq:wp}
\end{equation}
where $\Vert\bnabla\bm u\Vert=\sqrt{\sum_{i,j=1}^d (\partial_j u_i)^2}$ is the Frobenius norm of the velocity gradient with $d$ being the spatial dimension of the flow. 
The degree of compressibility satisfies $0\leqslant\wp\leqslant 1$, with $\wp=0$ for an incompressible
flow and $\wp=1$ for a potential flow $\bm u=\bnabla\phi$, where $\phi$ is a differentiable
scalar field. For free-surface turbulent flows---a case of particular relevance to our study---numerical simulations \citep{cdgs04,bdes04} and experiments \citep{cdgs04,coletti25,coletti_jfm} report values of $\wp$ between 0.34 and 0.5.

Lagrangian dynamics in compressible flows does not exhibit the universality of its incompressible counterpart. Specifically, the transport of scalar fields and tracers in fluctuating compressible flows has been found to depend strongly on the temporal correlation of the flow. For example, the inhomogeneity of the spatial distribution of tracer particles, in a compressible flow, can be enhanced or depleted by temporal correlations, depending on the spatial structure of the flow~\citep*{va97,bdes04,gm13,dmv14}. 
 
Stretching statistics in compressible flows and, in particular, the Cram\'er rate function has been studied in numerical simulations of free-surface turbulence~\citep*{bdl06} and two-dimensional chaotic flows~\citep{p-m14}. In general, $\lambda$ decreases with increasing $\wp$, i.e.~the mean rate of exponential stretching is reduced. For some random flows, there is even a critical degree of compressibility, $\wp_c$,
beyond which $\lambda$ becomes negative \citep{ckv98,gm13,dmv14}.
It is thus customary to distinguish between a weakly compressible regime ($0<\wp<\wp_c$) and a strongly compressible one ($\wp_c < \wp \leqslant 1$) \citep*{fgv01}. 
In the latter case, line elements contract in \textit{typical} realizations of the flow. 
However, because of the convexity of $L(q)$, high-order moments of $\ell(t)$ continue to grow in time, which indicates strong deviations from the mean contraction behaviour, i.e. rare events of intense stretching. 
 
The aforementioned relation between the stretching of small particles and that of material line elements implies that $\lambda$ determines the mean dynamics of a particle (stretching or contraction). However, the full statistics of the deformation, including the large deviations that can lead to breakup, are determined by $L(q)$. Hence, flows with the same $\lambda$ but different $L(q)$ yield different deformation statistics. This fact was demonstrated, in the incompressible case, by \citet*{ppv23} who compared the stretching statistics of polymers in isotropic turbulence to those in a Gaussian flow having the same $\lambda$ and the same correlation times of strain and vorticity. Small but systematic differences in the probability of large extensions were observed: stiff polymers stretched more in the turbulent flow (owing to extreme-straining events), while highly-elastic polymers stretched more in the Gaussian flow (owing to persistent mild straining). 

Here, we study how changes in $L(q)$, caused by flow compressibility, affect the statistics of particle deformation and breakup. 
Since the influence of compressibility on Lagrangian dynamics is flow-dependent, a universal relationship between particle stretching and flow compressibility is unlikely. 
We therefore study particular compressible flows and determine how the flow-specific changes in $L(q)$ impact particle stretching; we select random model flows for which $L(q)$ either is known analytically or may be computed accurately.

To gain insight into the effects of flow compressibility for a wide class of deformable bodies, we adopt
the vector model of 
\citet{orl82}, which generalizes the elastic dumbbell model to include the effects of internal viscosity and non-affine deformation.
This model was initially used to identify strong-flow conditions for spatially-uniform steady flows \citep{orl82}. In a series of subsequent papers, \citet*{swl91} and  \citet{sl92,sl93,sl94} used the same model to reveal the rich dynamics of orientation and stretching of deformable particles
 in inhomogeneous and unsteady flows. 
More recently, \citet{2023-cunha-etal} have used this model to test a new flow classification scheme. Obviously, a vector model is, in principle, only applicable to axisymmetric particles and is ill-suited to capture the high distortion that precedes a breakup event. Nevertheless, this minimal particle model, in conjunction with model random flows, provides an ideal setting for understanding the qualitative effects of flow compressibility on particle stretching and breakup.

The rest of this article is organized as follows. We first describe the vector model for a stretchable particle in \S~\ref{sec:micro_model}. Next, in \S~\ref{sec:st-dist}, we extend the theory of \citet{bfl00,bfl01} 
to general stretchable particles in compressible random flows and, thereby, derive a relationship between the stationary statistics of the particle size and $L(q)$.
In \S~\ref{sec:BK}, we  
present analytical results for the compressible
Batchelor--Kraichnan model, where the flow is Gaussian and has zero correlation time. In particular, we obtain the exact expression for the slope of the probability of intermediate particle sizes and calculate the mean time for breakups as a function of the degree of compressibility of the flow. An intriguing conclusion of the analysis is that stiff particles can breakup more rapidly in a compressible flow even as the mean straining is suppressed. 
 We then address the case of a random flow with nonzero correlation time in \S~\ref{sec:ren_flow},
 by studying 
 a compressible renewing flow,
 wherein the velocity gradient transforms randomly at fixed time instants. Importantly, the analytical predictions, obtained for the delta-correlated flow, anticipate all the qualitative features of the numerical results, computed for the time-correlated flow.
 We end in \S~\ref{sec:con} with a summary of our key findings and a discussion of their implications.

\section{Vector model of a stretchable particle}
\label{sec:micro_model}
The vector model of \citet{orl82} describes the size and orientation dynamics of a stretchable axisymmetric particle. The particle is assumed to be neutrally buoyant and smaller than the smallest length scale of the flow. In turbulent flows, this means that the size of the particle lies in the viscous dissipation range. Consequently, the centre of mass of the particle moves as material point, and the particle stretching and orientation dynamics is fully determined by the velocity gradient. The configuration of the particle is specified by a single
state vector $\bm R$, which provides its length $R=\vert\bm R\vert$ and orientation vector $\bm n=\bm R/R$. The
time evolution of $\bm R$ is governed by
\begin{equation}
    \dot{\bm R} 
    = 
    \osf\boldsymbol{\cdot}\bm R + g \left[\ssf\boldsymbol{\cdot}\bm R- \dfrac{\epsilon}{\epsilon+1}\,\dfrac{\bm R\boldsymbol{\cdot}\ssf\boldsymbol{\cdot}\bm R}{R^2}\,\bm R\right] -\dfrac{\bm R}{2\tau_p(\epsilon+1)},
    \label{eq:orl}
\end{equation}
where
$\osf$ and $\ssf$ are the antisymmetric and symmetric parts, respectively, of the velocity gradient at the position of the centre of mass of the particle.
Equation~\eqref{eq:orl} involves three parameters. The parameter $\tau_p$ is the relaxation time of the elastic restoring force that describes the resistance of the particle to stretching. 
The shape factor $g$, which may take values in the range $0\leqslant g\leqslant 1$, accounts for the non-affine deformation of particles that are not perfect rods: values of $g$ smaller than unity model the inefficiency of strain in rotating and stretching a particle of finite aspect ratio.
Finally, $\epsilon$ is the `internal viscosity' parameter and models internal dissipation mechanisms associated with the deformation of the particle.
It can assume non-negative values, ranging from zero to infinity; $\epsilon =0$ represents an elastically deformable particle, while the limit $\epsilon \rightarrow \infty$ models a rigid particle.  
Thus, \eqref{eq:orl} involves the interplay of rotational and extensional effects, induced by the fluid, as well as internal dissipation and relaxation of the particle. 
As described in \citet{orl82}, this equation encompasses several microscopic bodies for different values of the parameters,
including a linear elastic dumbbell ($\epsilon=0$, $g=1$,  $\tau_p>0$), a linear elastic dumbbell with internal viscosity ($\epsilon>0$, $g=1$,  $\tau_p>0$), a solid spheroidal particle ($\epsilon=\tau_p=\infty$, $0\leqslant g\leqslant 1$), and a material line element ($\epsilon=0$, $g=1$, $\tau_p=\infty$).
 
Equation~\eqref{eq:orl} can be re-written as coupled evolution equations for the size and orientation \citep{orl82}:
\begin{eqnarray}
    \dot{R}     &=& \dfrac{1}{ \epsilon +1} \bigg[ g \,  (\bm n\boldsymbol{\cdot}\ssf\boldsymbol{\cdot}\bm n)R - \frac{R}{2 \tau_p} \bigg],
    \label{eq:drdt}
    \\[2mm]
    \dot{\bm n} &=& \osf\boldsymbol{\cdot}\bm n + g \big[\ssf\boldsymbol{\cdot}\bm n -(\bm n\boldsymbol{\cdot}\ssf\boldsymbol{\cdot}\bm n)\bm n\big].
    \label{eq:dndt}
\end{eqnarray}
The latter equation is simply Jeffery's equation for the orientation vector of a rigid spheroidal particle. Furthermore, the orientation dynamics depends only on the parameter $g$, while the size dynamics depends on the two parameters $g/(\epsilon+1)$ and $1/2\tau_p (\epsilon+1)$. The limitations of this vector model become apparent on considering the case of $g=0$. While the equation for the orientation (\ref{eq:dndt}) reduces to that of a sphere, the equation for the extension (\ref{eq:drdt}) exhibits a nonphysical loss of flow-induced stretching (this breakdown of the model is not unexpected since an isotropic sphere cannot be described by a vector). Here, we are primarily concerned with elongated soft particles; in this context, the vector model provides a simple way to generalize beyond the elastic dumbbell and account for internal viscosity and the effects of a finite aspect ratio. 

The model of \citet{orl82} may be augmented in various ways. 
In \eqref{eq:orl}, the equilibrium length of the particle is zero, and therefore, $R$ vanishes in time in the absence of flow. A nonzero equilibrium length $\Req$ can be imposed either by modifying the form of the restoring force for small $R$ or by including Brownian fluctuations in \eqref{eq:orl}.
In addition, the restoring force is linear and thus allows infinite lengths. To introduce a maximum length $\Rmax$ in the model, it is sufficient to replace the linear force with a nonlinear one, such as that in the finite extensible nonlinear elastic (FENE) dumbbell model \citep{bird}. Furthermore, \eqref{eq:orl} does not preserve the volume of the particle. This problem can be rectified by allowing the parameters of the model to depend on $R$ in a suitable way \cite{ko86}.
As mentioned in the Introduction, our aim is to understand the qualitative effects of compressibility on particle stretching. 
Such results will be independent of the specific small-$R$ and large-$R$ regularizations of \eqref{eq:orl}.
For the derivation of our main predictions, we will therefore continue to use \eqref{eq:orl},
while interpreting the results within the range of particle sizes $\Req\ll R\ll\Rmax$.
A regularized version of the model will only be used, for practical reasons, in numerical simulations (see \S~\ref{sec:BK}) or in the explicit solution of the Batchelor--Kraichnan model (Appendix~\ref{sec:app-1}).

\section{Stationary size statistics in chaotic flows}
\label{sec:st-dist}

\subsection{The case of $\epsilon=0$ and $g=1$}

We first consider
the case $\epsilon=0$, $g=1$, $\tau_p>0$, i.e.~a Hookean dumbbell. 
We will then generalize the results to arbitrary values of the parameters in \S~\ref{subsec:general}.

\citet{bfl00,bfl01} studied the dynamics of a Hookean dumbbell in an incompressible flow, in the context of single-polymer dynamics in turbulence. In the absence of breakup events, they showed that
the stationary probability density function (PDF) of the size,
$P_{\rm{st}}(R)$,
displays a power-law behaviour:
\begin{equation}
    \label{eq:power}
    \pst(R)\sim R^{-1-\alpha}, \qquad \Req\ll R\ll\Rmax.
\end{equation}
The exponent $\alpha$ is determined by the rate function of the straining rate $\gamma(t)$. Given the large deviations form of the PDF of $\gamma(t)$ as
\begin{equation}
    P(\gamma,t)\sim e^{-S(\gamma-\lambda)t},
    \label{eq:pgamma}
\end{equation}
where $S$ is the rate function, one obtains $\alpha$ from 
\begin{equation}
\alpha=S'(\beta+\tau_p^{-1}-\lambda),
\end{equation}
where $S'$ denotes the derivative of $S$ and $\beta$ satisfies
\begin{equation}
    S(\beta+\tau_p^{-1}-\lambda)=\beta
    S'(\beta+\tau_p^{-1}-\lambda).
\end{equation}
(Here we follow the definition of \citealt{bfl00,bfl01}, where $S$ is such that $S(0)=0$. This definition is different from that used, for instance, in \citealt{cfvp91}, \citealt{bjpv98}, and \citealt{ccv2010}, where the rate function vanishes at $\lambda$. The two definitions only differ by a translation of the argument of the rate function.)

Alternatively, $\alpha$ can be expressed in terms of the generalized Lyapunov exponents \citep*{bcm03} as the solution of 
\begin{equation}
    \alpha=2\tau_pL(\alpha), 
    \label{eq:alpha_l}
\end{equation}
where the $q$th order generalized Lyapunov exponent $L(q)$ is defined as
\begin{equation}
\label{eq:gle}
    L(q) = \lim_{t \rightarrow \infty } \frac{1}{t} \log \Big\langle \brac{ \frac{ \ell(t)}{ \ell(0)}}^q \Big\rangle,
    \quad 
    \ell( t ) = |\bm{\ell} (t) |.
\end{equation}
In what follows, we shall use the latter formulation, wherein the connection between $\alpha$ and the statistics of the straining rate is more readily apparent. 

The preceding predictions hold for general random flows, the only assumption being that the correlation time of the velocity gradient, $\tau_f$, is finite
(in the limit where the Kubo number of the flow, $\mathit{Ku}=\lambda\tau_f$, tends to infinity, $\alpha$ approaches zero independently of the value of $\tau_p$; see \citealt{mv11}). In particular, though \citet{bfl00,bfl01} considered incompressible turbulence,
the theory developed therein holds for compressible flows as well. Indeed, the derivation of \eqref{eq:power} to \eqref{eq:alpha_l} relies on estimating the probability of strong stretching events, and, as mentioned in \S~\ref{sect:intro}, such events are also present in compressible chaotic flows.

\begin{figure}
    \centering
    \includegraphics[width=\textwidth]{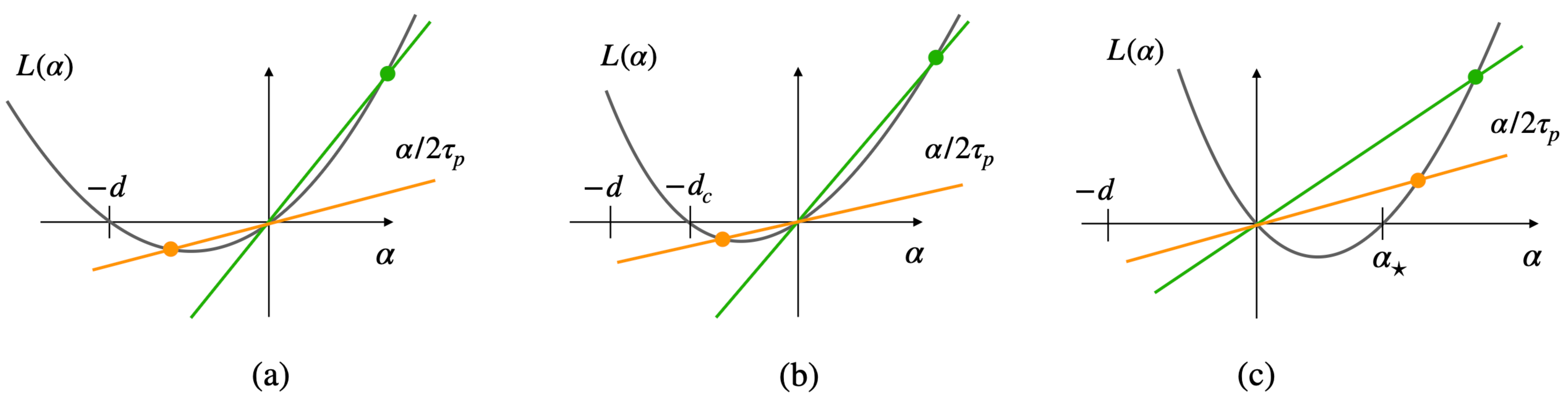}
    \caption{Graphical construction of the solution of \eqref{eq:alpha_l} in ({\it a}) the incompressible regime ($\wp=0$), ({\it b}) the weakly compressible regime ($0<\wp<\wp_c$), and ({\it c}) the strongly compressible regime ($\wp_c<\wp\leqslant 1$). The convex black curve is the graph of $L(\alpha)$, while the straight lines are $\alpha/(2\tau_p)$ for small (green line) and large (orange line) $\tau_p$.}
\label{fig:Lq_straight_line}
\end{figure}

The solution of \eqref{eq:alpha_l} depends on the form of $L(q)$ and therefore on the details of the random flow that transports the particle. Nevertheless, 
some aspects of the dependence of $\alpha$ on the system parameters can be deduced by using the general properties of $L(q)$, namely the convexity, the behaviour at the origin ($L(0)=0$ and $L'(0)=\lambda$), and the fact that the second zero of $L(q)$ is located at $q=-d_c\geqslant -d$ when $\lambda>0$.
For a given relaxation time $\tau_p$, \eqref{eq:alpha_l} has two solutions, which correspond to the two points of intersection
between the straight line $\alpha/2\tau_p$ and the graph of the function $L(\alpha)$
(see figure~\ref{fig:Lq_straight_line}).
One of the two points is the origin; the other gives the value of $\alpha$ that corresponds to $\tau_p$. 
Since the slope of the straight line $\alpha/2\tau_p$ decreases with $\tau_p$, 
an immediate consequence is that $\alpha$ is a decreasing function of $\tau_p$, i.e.~the probability of large sizes increases as the particle becomes more stretchable. Other features of $\alpha$ depend on the degree of compressibility of the flow.

We begin by recalling the  incompressible case ($\wp=0$, $\lambda>0$, and $d_c=d$) \citep{bfl00,bfl01}. 
The solution of this case is illustrated in figure~\ref{fig:Lq_straight_line}({\it a}).
For small values of $\tau_p$ (stiff particles), the straight line $\alpha/2\tau_p$ is steep, and its intersection with the graph of $L(\alpha)$ is located at a positive value of $\alpha$. Thus, $\pst(R)$ decays rapidly as $R$ deviates from the equilibrium size. As $\tau_p$ increases, the straight line $\alpha / {2 \tau_p }$ intersects the graph of ${L(\alpha)}$ at smaller and smaller values of $\alpha$. Consequently, $\pst(R)$ becomes less steep and the probability of large deformations increases, although, as long as $\alpha>0$, the configurations near the equilibrium one remain the most likely.
When $\tau_p$ equals $ ( 2 \lambda )^{-1}$ the slope of $\alpha/(2\tau_p)$ equals the slope of $L(\alpha)$ at the origin. Therefore, the two solutions of \eqref{eq:alpha_l} coincide, and $\alpha$ vanishes.
$\pst(R)$ now varies as $R^{-1}$ and is no longer normalizable in the limit $R_{\max} \rightarrow \infty $; in the absence of a nonlinear cutoff, the size of the majority of particles undergoes unbounded growth.
Following \citet{bfl00,bfl01},  the value $\tau_p=(2\lambda)^{-1}$ can therefore be regarded as the threshold for the transition from a shrunk state, close to equilibrium, to a highly stretched state. In the context of polymer dynamics, such a transition is known as
`the coil--stretch transition' \citep*{gcs05}; we shall here refer to it as the \textit{shrink--stretch} transition.

In the vicinity of the transition, $\alpha$ can be calculated exactly.
For small $q$, the function $L(q)$ is well-approximated by a quadratic polynomial, $L(q) \approx \lambda q + \Delta q^2 / 2$, where $\Delta$ is the variance of $\gamma(t)$ and is related to the rate function as $S'(0) = \Delta^{-1}$ \citep{cfvp91,bjpv98,ccv2010}. 
The quadratic behaviour of $L(q)$ near $q=0$ is consistent with the central limit theorem: the low-order moments of $\ell$ are indeed dominated by the typical fluctuations of the stretching rate. Plugging this approximation into \eqref{eq:alpha_l} yields \citep{bfl00,bfl01} 
    \begin{equation}
        \alpha = \frac{ 1 }{ \Delta } \brac{ \frac{1}{ \tau_p} - 2\lambda } .
        \label{eq:alpha_wi}
    \end{equation}
This expression holds for $\alpha$ close to zero irrespective of the statistics of the straining rate. 

As $\tau_p$ exceeds $( 2 \lambda )^{-1} $, the exponent $\alpha$ becomes negative,
which means that $\pst(R)$ is less steep than $R^{-1}$ and can even increase with $R$ (of course, nonlinear effects will eventually cause $\pst(R)$ to decrease rapidly as $R$ nears $\Rmax$).
The exponent saturates to $-d$ in the limit of $\tau_p\to\infty$: the straight line $\alpha/ 2\tau_p$ approaches the horizontal axis, in this limit, so that it intersects the graph of $L(\alpha)$ at $\alpha \to -d$. The scenario described thus far is well established in the context of polymer dynamics in incompressible turbulent flows \citep[see, e.g.,][]{wg10,bmpb12,ppv23}.

Let us now see how the scenario changes in the weakly compressible case, wherein $\wp>0$ and $\lambda$ decreases relative to its incompressible value while still remaining positive (hence, $d_c$ will also be positive but smaller than its incompressible value).
The solution of \eqref{eq:alpha_l} in the weakly compressible case is depicted in figure~\ref{fig:Lq_straight_line}({\it b}).
Applying the same arguments as in the incompressible case, we find that $\alpha$ is  positive
for $\tau_p < ( 2 \lambda )^{-1}$, vanishes for $\tau_p = ( 2 \lambda )^{-1}$, takes negative values $-d_c<\alpha<0$ for $\tau_p > ( 2 \lambda )^{-1}$, and tends to $-d_c$ as $\tau_p\to\infty$. 
It is not possible to make general statements regarding the dependence of $\alpha$ on $\wp$ for small values of $\tau_p$, because this relationship is sensitive to the specific form of $L(q)$.
However, since $\lambda$ and $d_c$ generally decrease with $\wp$, we expect the shrink--stretch transition to shift to larger values of $\tau_p$. And, for large $\tau_p$, we expect the probability of large deformations  to be subdued by compressibility (indeed $\alpha\geqslant -d_c>-d$).

In a strongly compressible flow ($\wp>0$ and $\lambda<0$),
the statistics of the particle size differs significantly from the two previous cases.
Since $L'(0)=\lambda<0$, the function $L(q)$ is positive for all $q<0$. Consequently, $\alpha$ cannot become negative. Its value is actually bounded below by the positive zero of $L(q)$, i.e. $\alpha\geqslant\alpha_\star$, where $\alpha_\star$ is such that $L(\alpha_\star)=0$ and $\alpha_\star>0$ (see figure~\ref{fig:Lq_straight_line}{\it c}).
The shrink--stretch transition is therefore not observed in a strongly compressible flow and, at large $\tau_p$, the probability of strong deformations is depleted by compressibility.
It is interesting to note, however, that the PDF of the particle size continues to behave as a power law, and therefore significant deviations from the equilibrium size remain possible, despite the fact that in typical realizations of the flow particles contract ($\lambda<0$).

\subsection{The general case}

\label{subsec:general}

Up till now, we have examined the special case where the shape factor $g$ is unity and the internal viscosity coefficient $\epsilon$ is zero, which corresponds to a Hookean dumbbell. The behaviour of the exponent $\alpha$ for $0 < g < 1$ and $\epsilon > 0$ can be obtained from the $g=1$, $\epsilon=0$ case by using a scaling argument, as in \citet{v21}.
From \eqref{eq:drdt}, we deduce that the size dynamics of a particle with $0 < g < 1$ and $\epsilon > 0$ is equivalent to that of a Hookean dumbbell having
relaxation time $(1+\epsilon)\tau_p$ in a flow with rate-of-strain tensor $ g \ssf / (1+\epsilon)$. 
To understand the effect of rescaling the rate-of-strain tensor $\ssf$, we return to the time evolution of an infinitesimal line element along a trajectory $\bm x(t)$: 
\begin{equation}
 \label{eq:dtl}   \dfrac{\mathrm{d}\ell_i}{\mathrm{d}t} = \ell_j\,\partial_j u_i(\bm x(t),t). 
\end{equation}
The length of the line element evolves according to the equation
\begin{equation}\label{eq:ell}
    \dfrac{\mathrm{d}}{\mathrm{d}t}\ln \ell =  \hat{\bm\ell} \boldsymbol{\cdot} \ssf  \boldsymbol{\cdot}  \hat{\bm\ell},
\end{equation}
whose formal solution is 
\begin{equation}
\ell(t) = \ell(0) \exp \! \left[ \int_0^t \mathrm{d}s \,(\bm{\hat{ \ell}} \boldsymbol{\cdot} { \ssf } \boldsymbol{\cdot}  \bm{\hat{ \ell}}) \right].
\label{eq:l-grad}
\end{equation}
Consider now 
the rescaled rate-of-strain tensor $\ssf_\xi=\xi\ssf$ and denote the $q$th generalized Lyapunov exponent associated with $\ssf_\xi$ as $L_\xi(q)$.
From \eqref{eq:l-grad} and the definition of the generalized Lyapunov exponents in \eqref{eq:gle}, it is clear that 
$L_\xi(q) = L(\xi q)$, i.e.~a rescaling of the rate-of-strain tensor results in a rescaling of the argument of $L(q)$ by the same factor. 
Therefore, \eqref{eq:alpha_l} yields
\begin{equation}
    \alpha( \epsilon, g, \tau_p ) = 2 (1+\epsilon) \tau_p \,L_{\frac{g}{1+\epsilon}}\brac{\alpha( \epsilon, g, \tau_p ) }
    = 2 (1+\epsilon) \tau_p\,L\brac{ \frac{g}{1+\epsilon}\,\alpha( \epsilon, g, \tau_p ) }
\end{equation}
or, equivalently, 
\begin{equation}
    \dfrac{g}{1+\epsilon}\:\alpha( \epsilon, g, \tau_p ) 
    = 2 g \tau_p\,L\brac{ \frac{g}{1+\epsilon}\,\alpha( \epsilon, g, \tau_p ) }.
\end{equation}
By comparing the latter equation with \eqref{eq:alpha_l}, we find
\begin{equation}
    \alpha( \epsilon, g, \tau_p ) = \frac{ 1+\epsilon }{g}\: \alpha( 0, 1, g\tau_p ).
    \label{eq:alpha_scaling}
\end{equation}
The dependence of $\alpha$ on the shape factor and the internal viscosity  is encapsulated in this relation. In particular, internal viscosity increases the absolute value of $\alpha$ and therefore steepens the PDF of the particle size at both small $\tau_p$, where $\alpha$ is positive, and large $\tau_p$, where $\alpha<-1$, thus making the shrink--stretch transition sharper \citep{v21}. 
Reducing $g$ has the effect of increasing $\alpha$, both because the prefactor increases and because
the effective relaxation time decreases (recall that $\alpha$ is a decreasing function of $\tau_p$). 
At small $\tau_p$, this results in a depletion of the tail of $\pst(R)$, consistent with the interpretation of $g$ as a measure of the efficiency of strain in stretching particles. At large $\tau_p$, 
the effect is more subtle and depends on the details of how $\alpha$ varies with $\tau_p$. Indeed, as long as $\tau_p$ is such that $\alpha(0,1,\tau_p)<0$, 
decreasing $g$ amplifies the prefactor $1/g$, but multiplying the relaxation time by $g<1$ reduces the absolute value of $\alpha$. 

Thanks to \eqref{eq:alpha_scaling}, we can understand the influences of internal viscosity and the shape factor using just the $\epsilon=0$, $g=1$ case. So, we shall focus on this case in our subsequent investigation of the effects of compressibility. The general properties of $L(q)$ have already yielded some qualitative insights into the dependence of $\alpha$ on $\wp$, especially for large $\tau_p$. For a more detailed understanding, we shall study specific examples of compressible random flows: the next section discusses the Batchelor--Kraichnan flow, which admits an analytical solution, while \S~\ref{sec:ren_flow} presents numerical simulations for a renewing flow.
We shall see that compressibility can have counterintuitive effects, particularly for the breaking of stiff particles.

\section{Delta-correlated Batchelor--Kraichnan flow}
\label{sec:BK}

\subsection{The compressible Batchelor--Kraichnan model}

In the Batchelor regime of the Kraichnan model, the velocity field is Gaussian, has zero correlation time, and is statistically homogeneous, isotropic, parity invariant, and linear in space \citep{fgv01}. 
The second-order correlation of the
velocity components is 
\begin{equation}
    \langle u_i(\bm x,t)u_j(\bm y,t')\rangle 
    = \{ 2D_0\delta_{ij}-D[(d+1- 2 \wp)r^2\delta_{ij} - 2(1-d\wp) r_ir_j] \}\delta(t-t'),
\end{equation}
where $i,j=1,\dots,d$,
the separation vector is defined as $\bm r=\bm x-\bm y$, $D_0$ and $D$ are positive constants (in particular, $D$ has the dimension of inverse time), and the degree of compressibility $\wp$ has been defined in \eqref{eq:wp}.  
The corresponding \textit{Eulerian} correlation of the velocity gradient is \citep{fgv01}
\begin{equation}
    \label{eq:corr-grad}
    \langle \partial_j u_i (\bm x,t)\partial_l u_k(\bm x,t')\rangle = 2\mathsfi{C}_{ijkl}\,\delta(t-t'),
    \qquad
    i,j,k,l=1,\dots,d,
\end{equation}
where 
\begin{equation}
\label{eq:C-tensor}
    \mathsfi{C}_{ijkl} = D [(d+1-2\wp)\delta_{ik}\delta_{jl}+(\wp d-1)(\delta_{ij}\delta_{kl}+\delta_{il}\delta_{jk})].
\end{equation}
The \textit{Lagrangian} correlation of the velocity gradient, relevant for line-element stretching, is the same as the Eulerian one because of the $\delta$-correlation in time of the Batchelor--Kraichnan model.

Now, to study the evolution of a line element, it is convenient to replace the Lagrangian velocity gradient $\bnabla\bm u(\bm x(t),t)$ in \eqref{eq:dtl} with a random tensor having the same statistics, i.e. to consider the stochastic differential equation
\begin{equation}
\label{eq:lG}
\dfrac{\mathrm{d}\bm \ell}{\mathrm{d}t}=\G(t)\bm\ell,
\end{equation}
where $\G(t)$ is a Gaussian $d\times d$ tensorial noise such that $\langle\G(t)\rangle=0$ and
\begin{equation}
\label{eq:corr-G}
\langle\mathsfi{G}_{ij}(t)\mathsfi{G}_{kl}(t')\rangle=2\mathsfi{C}_{ijkl}\,\delta(t-t').
\end{equation}
It is important to note, however, that while
the It\^o and Stratonovich interpretations of \eqref{eq:dtl} coincide, \eqref{eq:lG} and \eqref{eq:dtl} are statistically equivalent only if the former is interpreted in the It\^o sense (see Gaw\c{e}dzki \citeyear{g08}, \S\S~2.4.2 and~2.9.3, for a detailed discussion).
This proviso is particularly relevant in the compressible case, wherein the It\^o and Stratonovich interpretations of \eqref{eq:lG} differ, with 
the Stratonovich counterpart of \eqref{eq:lG} being
\begin{equation}
\label{eq:Stratonovich}
\dfrac{\mathrm{d}\bm \ell}{\mathrm{d}t}=\G(t)\circ\bm\ell - \widetilde{\mathsfbi{C}} \,\bm\ell,
\end{equation}
where $\widetilde{\mathsfi{C}}_{il}=\mathsfi{C}_{ijjl}=\wp D(d-1)(d+2)\delta_{il}$
(clearly, the additional drift term vanishes if the flow is incompressible). The difference between the two interpretations becomes important when trying to approach the delta-correlated limit by decreasing the correlation time of a time-correlated random flow---such a limiting procedure yields the Stratonovich interpretation provided the correlation function of the nonwhite noise is time-reversible~\citep{Celani2013-colored-noise}. In \S~\ref{sec:ren_flow}, we shall compare the stretching statistics in the Batchelor--Kraichnan flow with that in a time-correlated renewing flow; in order for the statistics in the two flows to agree, in the small correlation-time limit, we will have to account for the drift term of \eqref{eq:Stratonovich} while prescribing the Lagrangian velocity gradient of the renewing flow.

We come now to the generalized Lyapunov exponents of the compressible Batchelor--Kraichnan flow. As a consequence of the Gaussian statistics and the temporal decorrelation of the velocity field, $L(q)$ is quadratic for all $q$ and not just near $q=0$ \citep{ckv98,fgv01}:
\begin{equation}
L(q)=\lambda q + \dfrac{\Delta}{2}\, q^2
\end{equation}
with 
\begin{equation}
    \lambda = D(d-1)(d-4\wp), 
    \quad 
    \Delta=2D(d-1)(1+2\wp).
    \label{eq:lam_del}
\end{equation}
(In the above expressions, $d>1$; the case $d=1$ is recovered from the same formula by taking $D\propto 1/(d-1)$; see \citealt{fgv01}.)
It follows that: (i) the critical degree of compressibility, above which $\lambda<0$, is $\wp_c=d/4$; since $\wp\leqslant 1$, the strongly compressible regime only exists for $d\leqslant 4$; (ii) $L(q)$ vanishes at $q=0$ and $q=-{2\lambda} /{\Delta}=-\brac{d-4\wp}/ \brac{1+2\wp}$; therefore, for $\wp< d/4$, the correlation dimension is  $d_c=\brac{d-4\wp}/ \brac{1+2\wp}$;  (iii) $L(2)$ is independent of $\wp$ for all $d$, i.e.~the growth rate of the mean square of the length of a line element is independent of the degree of compressibility. 

\begin{figure}
 \centering
         \includegraphics[width=0.42\textwidth]{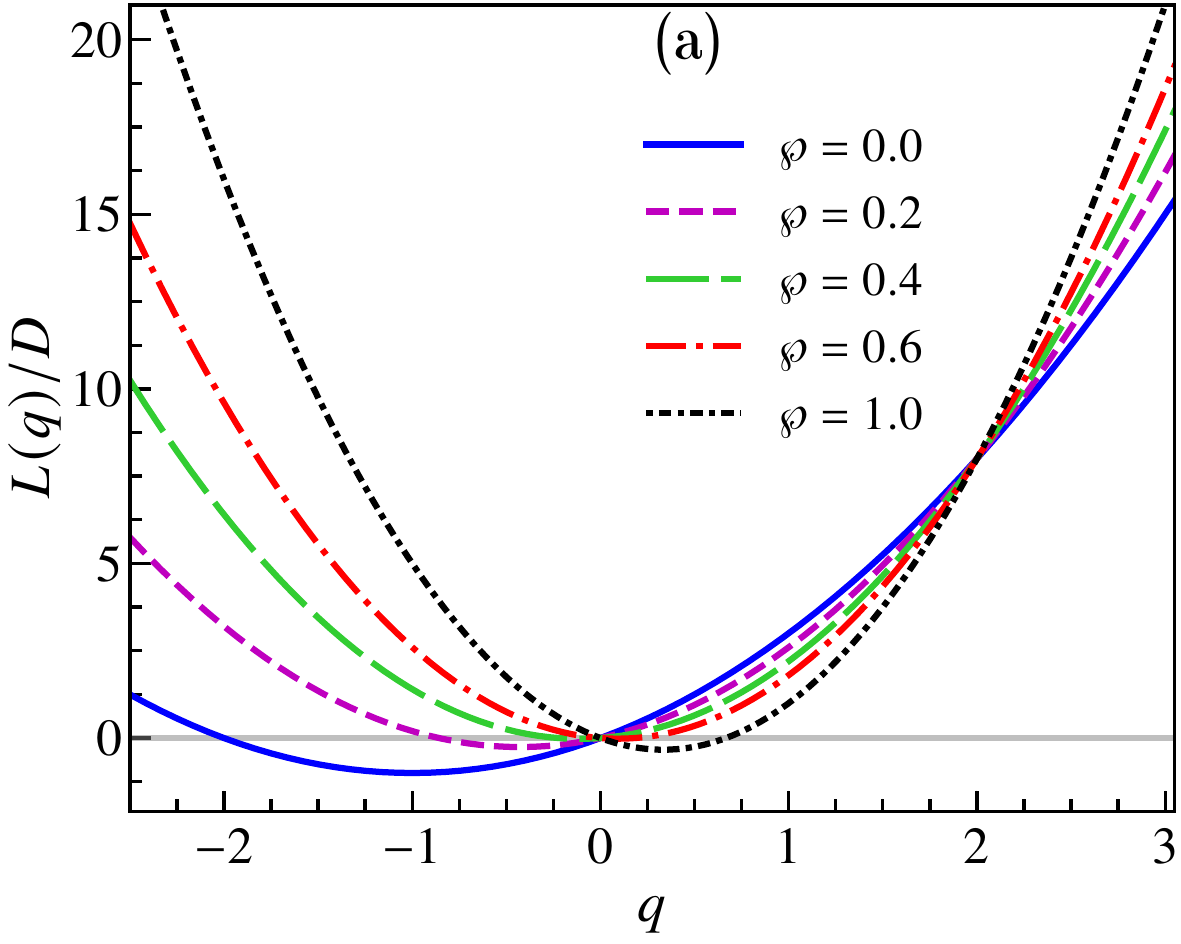}\\
         \includegraphics[width=0.42\textwidth]{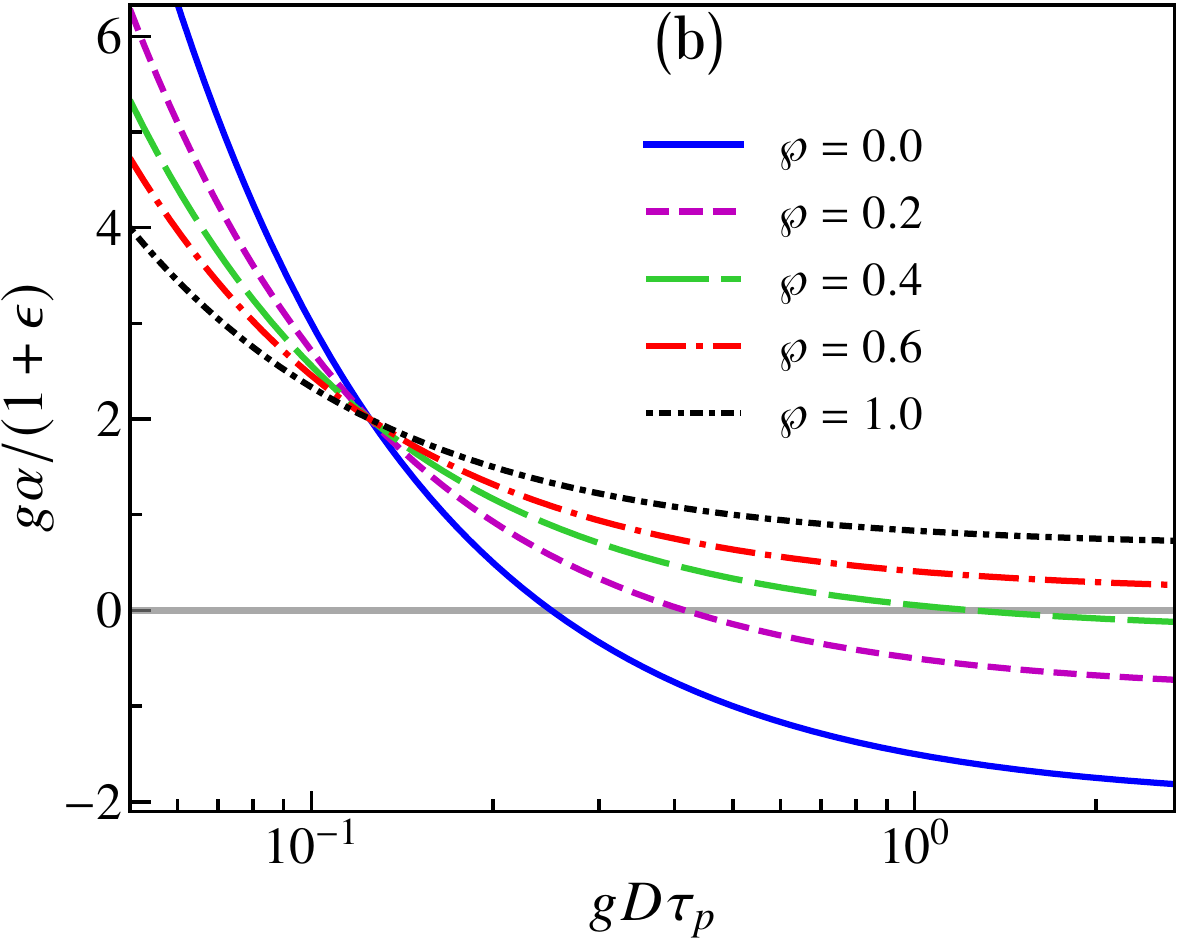}\hspace{2 em}
         \includegraphics[width=0.42\textwidth]{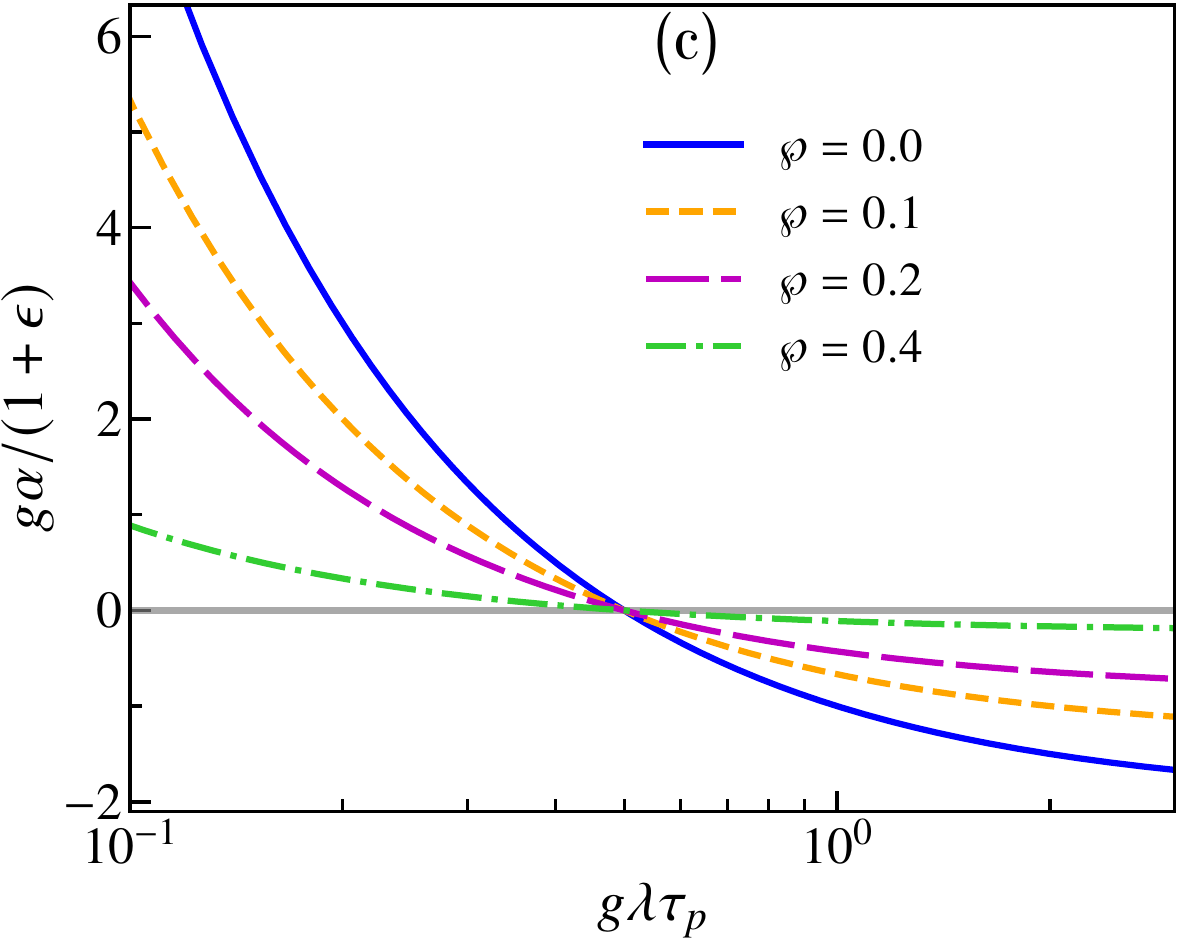}
        \caption{Plots of ({\it a}) $L(q)/D$ as a function of $q$, ({\it b}) $g \alpha /(1+\epsilon)$ as a function of $gD\tau_p$, and ({\it c}) $g \alpha /(1+\epsilon)$ as a function of $g \lambda \tau_p$, for different values of $\wp$ in the Batchelor--Kraichnan flow. In all panels, $d=2$. The values of $\wp$ considered in ({\it a}) and ({\it b}) cover three regimes of compressibility: incompressible ($\wp=0$), weakly compressible ($0<\wp < d/4$), and strongly compressible $(d/4 < \wp \leqslant 1)$;  only the incompressible and weakly compressible regimes are considered in ({\it c}).}
        \label{fig:bk-plots}
    \end{figure}

The function $L(q)$ is plotted in figure~\ref{fig:bk-plots}({\it a}) for $d=2$ and various values of $\wp$ (the qualitative behaviour of $L(q)$ is similar for all $d < 4$).
Two properties of $L(q)$ are worth highlighting.
First, even in the strongly compressible regime, where $\lambda<0$, the moments $\langle\ell^q(t)\rangle$ grow in time as soon as $q> \brac{4 \wp - d} / \brac{1+2 \wp}$, where the maximum value of the lower bound is $(4-d)/3$ (for $\wp=1$). This growth in time of relatively low order moments of $\ell$ implies that line elements experience episodes of strong stretching, which deviate significantly from the mean shrinking behaviour.
Second, $L(q)$ increases with $\wp$ for $q\geqslant 2$, i.e.~
strong-stretching events, caused by extreme fluctuations of the the strain-rate ($\hat{\bm\ell} \boldsymbol{\cdot} \ssf  \boldsymbol{\cdot}  \hat{\bm\ell}$ in \eqref{eq:ell}),
not only exist in the compressible regime, but become more prominent with increasing compressibility.
Next, we examine the consequences of these features of $L(q)$ on the statistics of particle stretching, in particular, on the exponent $\alpha$ that determines the power-law form of the stationary PDF of the particle size.

\subsection{Stationary PDF of the particle size}

\label{subsec:stationary_BK}

Since $L(q)$ is quadratic in $q$, the exponent $\alpha$ is obtained by inserting  $\lambda$ and $\Delta$ from \eqref{eq:lam_del} into \eqref{eq:alpha_wi} and \eqref{eq:alpha_scaling}:
\begin{equation}
\label{eq:alpha_BK}
    \alpha = \frac{1+\epsilon}{g(1+2\wp)} \left[ \frac{1}{2(d-1)gD\tau_p} - d+4\wp\right].
\end{equation}
(In Appendix~\ref{sec:app-1}, $\alpha$ is calculated in alternative way by solving the Fokker--Planck equation for the PDF of $\bm R(t)$.)
The rescaled exponent is plotted as a function of $gD\tau_p$, in figure~\ref{fig:bk-plots}({\it b}),
for various compressibilities and for $d=2$ (the behaviour is analogous for any $d<4$).
In the incompressible and weakly compressible cases ($0\leqslant \wp<\wp_c$), $\alpha$ decreases from positive to negative values as $gD\tau_p$ increases (vanishing at $g\tau_p=(2\lambda)^{-1}$, which marks the shrink--stretch transition). In contrast, $\alpha$ is always positive in the strongly compressible case ($\wp_c<\wp\leqslant 1$). 
The curves $\alpha$ vs $gD\tau_p$ cross at $gD\tau_p=[2(d-1)(d+2)]^{-1}$. Therefore, two different situations can be identified.
At large $gD\tau_p> [2(d-1)(d+2)]^{-1}$, the effect of compressibility is to increase $\alpha$; in particular, the asymptotic value of $\alpha$ is
\begin{equation}
\lim_{D\tau_p\to\infty}\alpha = -\dfrac{(1+\epsilon)(d-4\wp)}{g(1+2\wp)}.
\end{equation}
These features of the variation of $\alpha$ with $\wp$ and $\tau_p$ are entirely consistent with the predictions, of \S~\ref{sec:st-dist}, that follow from the general properties of $L(q)$. The behaviour of $\alpha$ at small $\tau_p$, however, depends on the flow-specific behaviour of $L(q)$ at large $q$.  
Figure~\ref{fig:bk-plots}({\it b}) shows that, for small $gD\tau_p < [2(d-1)(d+2)]^{-1}$, compressibility decreases the value of $\alpha$. This can be confirmed by comparing $\alpha$ with its value for $\wp=0$ (denoted as $\alpha_0$) in the limit of vanishing relaxation times:
\begin{equation}
\lim_{D\tau_p\to 0}\dfrac{\alpha}{\alpha_0} = \dfrac{1}{1+2\wp}.
\end{equation}
As noted earlier, compressibility simultaneously decreases $\lambda$ and increases the probability of large fluctuations of the line-element stretching rate ($L(q)$ increases with $\wp$ for $q>2$).
To isolate the consequences of the latter effect on $\alpha$, figure~\ref{fig:bk-plots}({\it c}) presents the rescaled exponent as a function of $g \lambda \tau_p$ (in the incompressible and weakly compressible regimes). This plot accounts for the displacement of the zero of $\alpha$ due to the variation of $\lambda$ with $\wp$ (all curves now pass through zero at the same point); the large decrease of $\alpha$ with compressibility, seen in figure~\ref{fig:bk-plots}({\it c}) for small $\tau_p$, is entirely caused by strong fluctuations of the strain rate.

 The above results imply that flow compressibility displaces the shrink--stretch transition to higher values of $D\tau_p$ and eventually suppresses the transition once the strongly compressible regime is attained. Moreover, compressibility depletes the probability of large extensions for highly elastic particles (large $D\tau_p$) and increases it for stiff particles (small $D\tau_p$). Consequently, the shrink--stretch transition becomes evermore gradual as $\wp$ increases, until it disappears completely in the strongly compressible regime.
The contrasting effects of compressibility on elastic and stiff particles arise from the differing response of these particles to fluctuations of the strain rate ($\bm n\boldsymbol{\cdot}\ssf\boldsymbol{\cdot}\bm n$ in \eqref{eq:drdt}, see also \eqref{eq:ell}).

Elastic particles integrate the straining action of the rate-of-strain tensor $\ssf$ over long times; therefore, the large but rare fluctuations of the strain rate are filtered out, and the particle dynamics is dominated by mild and frequent fluctuations, which become weaker as compressibility increases.
Stiff particles have a very short response time to the fluctuations of the strain rate; their dynamics is therefore sensitive to large fluctuations even if these are short-lived, and the frequency of large fluctuations of the strain rate increases with $\wp$. Therefore, compressibility increases the probability of highly-stretched stiff particles.

\subsection{Mean breakup time}

We have thus far analyzed the stationary PDF of particle size while ignoring the possibility of breakup under the action of strong velocity gradients. We now
assume that a particle breaks when the elastic force in \eqref{eq:orl} exceeds a threshold, which is equivalent to assuming that breakup occurs when the particle size exceeds a threshold extension $\Rbr$ (with $\Req\ll\Rbr\ll\Rmax$). We aim to study the mean breakup time $T_{\text{br}}$ and its dependence on flow compressibility. 

As shown in Appendix~\ref{sec:app-1}, the PDF of $R$ at time $t$, denoted as $P(R,t)$, satisfies the following Fokker--Planck equation:
\begin{equation}
    \partial_t P = - \partial_R  [\mathfrak{D}_1(R) P ] + \partial_R^2 [\mathfrak{D}_2(R) P],     \label{eq:FPE}
\end{equation}
where the expressions for the drift and diffusion coefficients are given in Appendix~\ref{sec:app-1} for $\epsilon=0$ and $g=1$. Here, we obtain asymptotic results for the general case, using just the power-law behaviour of the stationary PDF for $\Req\ll R\ll\Rmax$ and the fact that $\mathfrak{D}_2\sim R^2$ in the same range of extensions (the diffusion coefficient must scale as $R^2$ in view of the linearity of the velocity field in the neighbourhood of the particle).

In the absence of breakups, \eqref{eq:FPE} would be complemented by reflecting boundary conditions at $R=0$ and $R=\Rmax$ (i.e.~the probability current $\partial_{R} (\mathfrak{D}_2 P) - \mathfrak{D}_1 P$ is set to zero at both ends of the interval $0\leqslant R\leqslant\Rmax$). The corresponding steady solution is the stationary PDF
$\pst(R)\propto e^{-\Phi_a(R)}$, where
\begin{equation}
\label{eq:Phi}
\Phi_a(R) = \ln \mathfrak{D}_2(R) - \int_a^{R} \mathrm{d}z\, \frac{\mathfrak{D}_1(z)}{\mathfrak{D}_2(z)}
\end{equation}
and the value of the positive constant $a$ is fixed by the normalization of $\pst(R)$ \citep{gardiner}. Appendix~\ref{sec:app-1} shows that $\pst(R)\sim R^{-1-\alpha}$ for $\Req \ll R \ll \Rmax$ (with $\alpha$ given by \eqref{eq:alpha_BK}), in agreement with the general theory of \S~\ref{sec:st-dist}.

Now, to account for particle breakups, we impose an absorbing boundary condition on the right-hand side, at the threshold value $\Rbr$, i.e.~we set $P(\Rbr,t)=0$ for all $t$.  
Given a particle with initial size $R(0)=\rho$, such that $\Req\leqslant\rho<\Rbr$,
we obtain the average time it takes to break, $\Tbr$, by calculating the mean first-passage time through the boundary at $R=\Rbr$. We thus have (see \citealt{gardiner}, \S~5.2.7)
\begin{equation}
    \label{eq:Tbr}
\Tbr =
\int_\rho^{\Rbr}
\mathrm{d}y
\int_0^{y}
\dfrac{\mathrm{d}z}{\mathfrak{D}_2(z)}\,
\exp\left[
-\int_z^y \mathrm{d}x\,\dfrac{\mathfrak{D}_1(x)}{\mathfrak{D}_2(x)}
\right].
\end{equation}
(We have rewritten the original expression in \citealt{gardiner} so as to avoid the divergence of $\Phi_a(R)$ for $a\to 0$.)

We can estimate the asymptotic behaviour of $\Tbr$ as a function of $\Rbr/\rho$, in the limit of $\Rbr\gg\Req$, by ignoring the contribution of the range $0\leqslant z\leqslant \Req$
to the second integral in \eqref{eq:Tbr}; this amounts to ignoring the small amount of time spent by the particle near $\Req$ compared to the time it takes to reach $\Rbr$.
Since $e^{-\Phi_a(R)}\sim R^{-1-\alpha}$  and $\mathfrak{D}_2(R)\sim R^2$ for $\Req\ll R\ll\Rmax$, we have
\begin{equation}
   \exp\left[
-\int_z^y \mathrm{d}x\,\dfrac{\mathfrak{D}_1(x)}{\mathfrak{D}_2(x)}
\right] \sim \left(\dfrac{z}{y}\right)^{1-\alpha}.
\end{equation}
On inserting the latter expression into \eqref{eq:Tbr}, we find
\begin{equation}
\label{eq:Tbr_asymptotic}
    D\Tbr \sim
\operatorname{sgn}(\alpha)\int_\rho^{\Rbr}
\mathrm{d}y\, 
 y^{\alpha-1}(\Req^{-\alpha}-y^{-\alpha})
\propto
\begin{cases}
(\Rbr/\rho)^\alpha, & \alpha>0, 
\\
\ln(\Rbr/\rho), & \alpha<0,   
\end{cases}
\end{equation}

This result implies that,
in the incompressible and weakly compressible regimes ($0\leqslant\wp<\wp_c$), the mean breakup time increases as $(\Rbr/\rho)^\alpha$ for $\tau_p<(2\lambda)^{-1}$ and as the logarithm of $\Rbr/\rho$ for  $\tau_p>(2\lambda)^{-1}$. In the strongly compressible regime ($\wp_c<\wp\leqslant 1)$, the breakup time exhibits a power-law variation with $\Rbr$ for all $\tau_p$.

The dependence of $\Tbr$ on $\epsilon$ and $g$ cannot be deduced entirely from \eqref{eq:Tbr_asymptotic}. 
However, we can use
the scaling relation in \eqref{eq:alpha_scaling} to predict that internal viscosity increases the slope of the power law when $\alpha>0$, consistent with the fact that internal viscosity opposes sudden size changes that could result in particle breakups.
The shape factor can have a more significant effect: 
unlike $\epsilon$, a variation of the shape factor can cause a change in the sign of $\alpha$, leading to a transition between a logarithmic and a power-law growth of $\Tbr$. 

\begin{figure}
    \centering
    \includegraphics[width=0.42\textwidth]{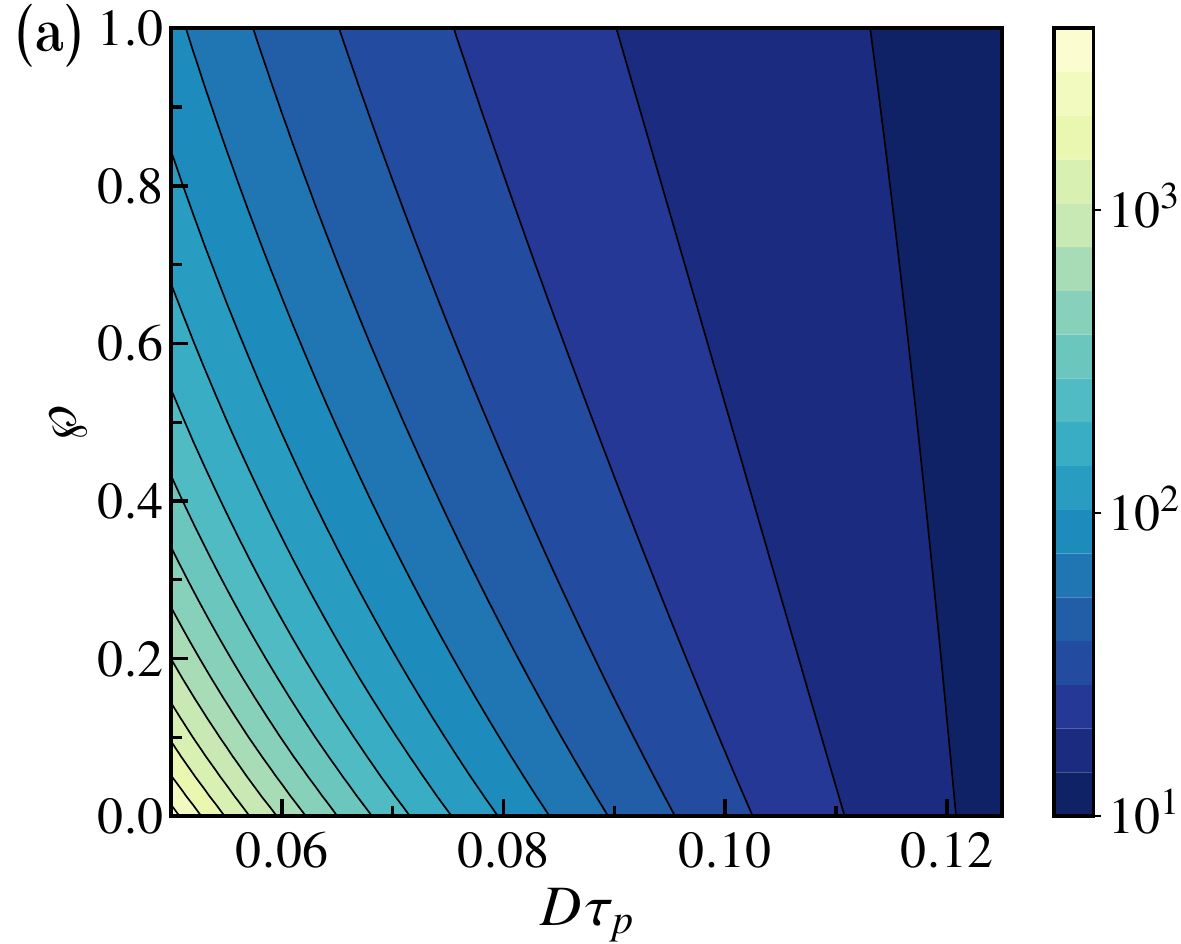}\hspace{2 em}%
    \includegraphics[width=0.42\textwidth]{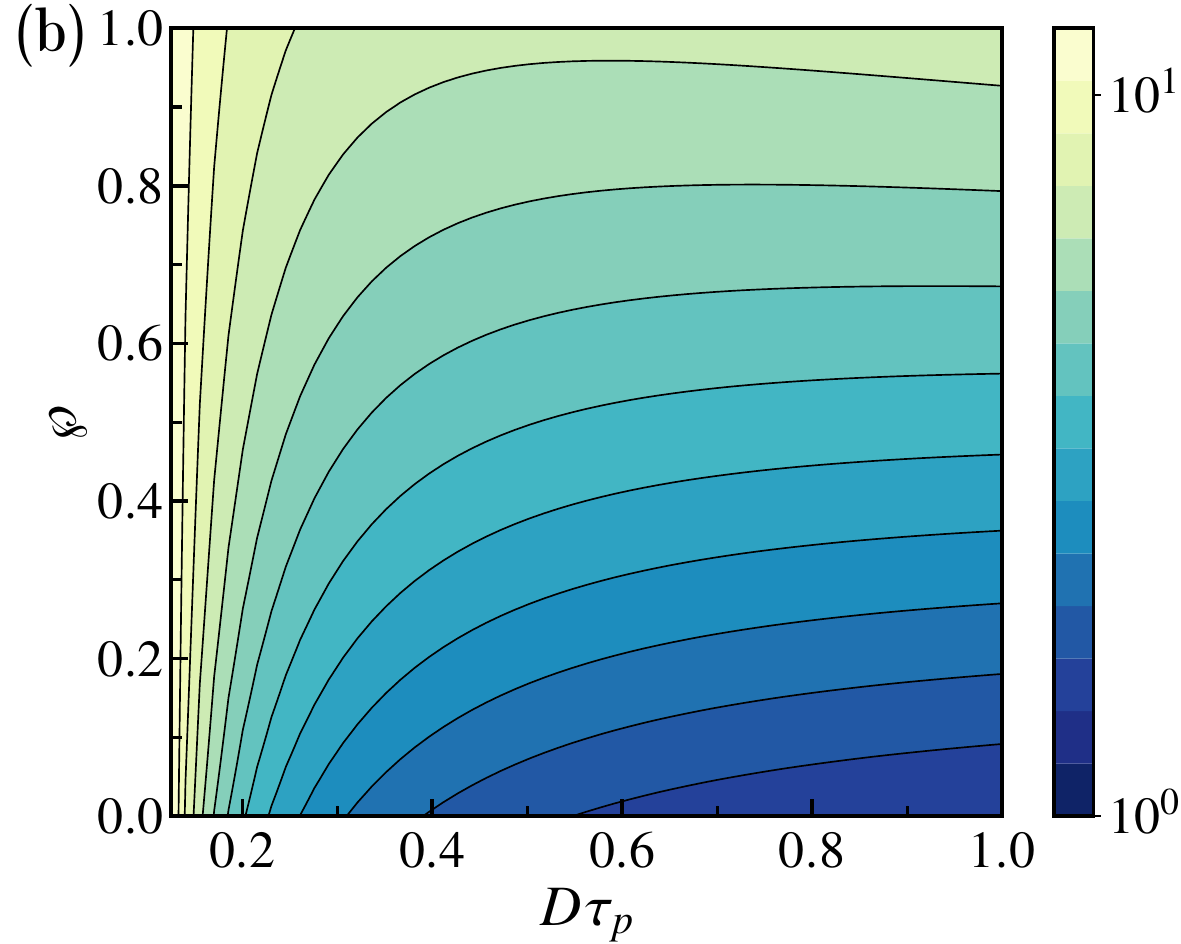}
    \caption{Contour plot of the mean breakup time $\Tbr$ as function of $D\tau_p$ and $\wp$ (note the logarithmically spaced contours) for ({\it a}) small and ({\it b}) large values of $D\tau_p$. In both panels, $d=2$, $\epsilon=0$, $g=1$, $\Rbr/\rho=10$, and the elastic force is linear.}    \label{fig:Tbr}
\end{figure}

The dependence of $\Tbr$ on the degree of compressibility is more difficult to estimate analytically. We therefore insert the expressions of $\mathfrak{D}_1(R)$ and $\mathfrak{D}_2(R)$ for a Hookean dumbbell, given in \eqref{eq:D1_D2_BK}, into \eqref{eq:Tbr} and calculate $\Tbr$ numerically (since $\Rbr\ll\Rmax$, we have used a linear elastic force, i.e.~$f(R)=1$ in \eqref{eq:D1_D2_BK}).
The variation of $\Tbr$ with $D\tau_p$ and $\wp$ is depicted in figures~\ref{fig:Tbr}({\it a,b}), for small and large $D\tau_p$, respectively.
Unsurprisingly, the breakup time decreases as the particle becomes more elastic, i.e. $\Tbr$ is a decreasing function of $D\tau_p$. With regard to compressibility, its contrasting effects on the stretching of stiff and elastic particles, uncovered in \S~\ref{subsec:stationary_BK}, naturally produce contrasting effects on the breakup time:
$\Tbr$ decreases with $\wp$ at small $D\tau_p$ (precisely, for $D\tau_p<[2(d-1)(d+2)]^{-1}=1/8$)
and increases with $\wp$ at large $D\tau_p$. So, stiff particles not only stretch more but also breakup faster in compressible flows. 

\section{Time-correlated renewing flow}
\label{sec:ren_flow}
\begin{figure}
 \centering
         \includegraphics[width=0.42\textwidth]{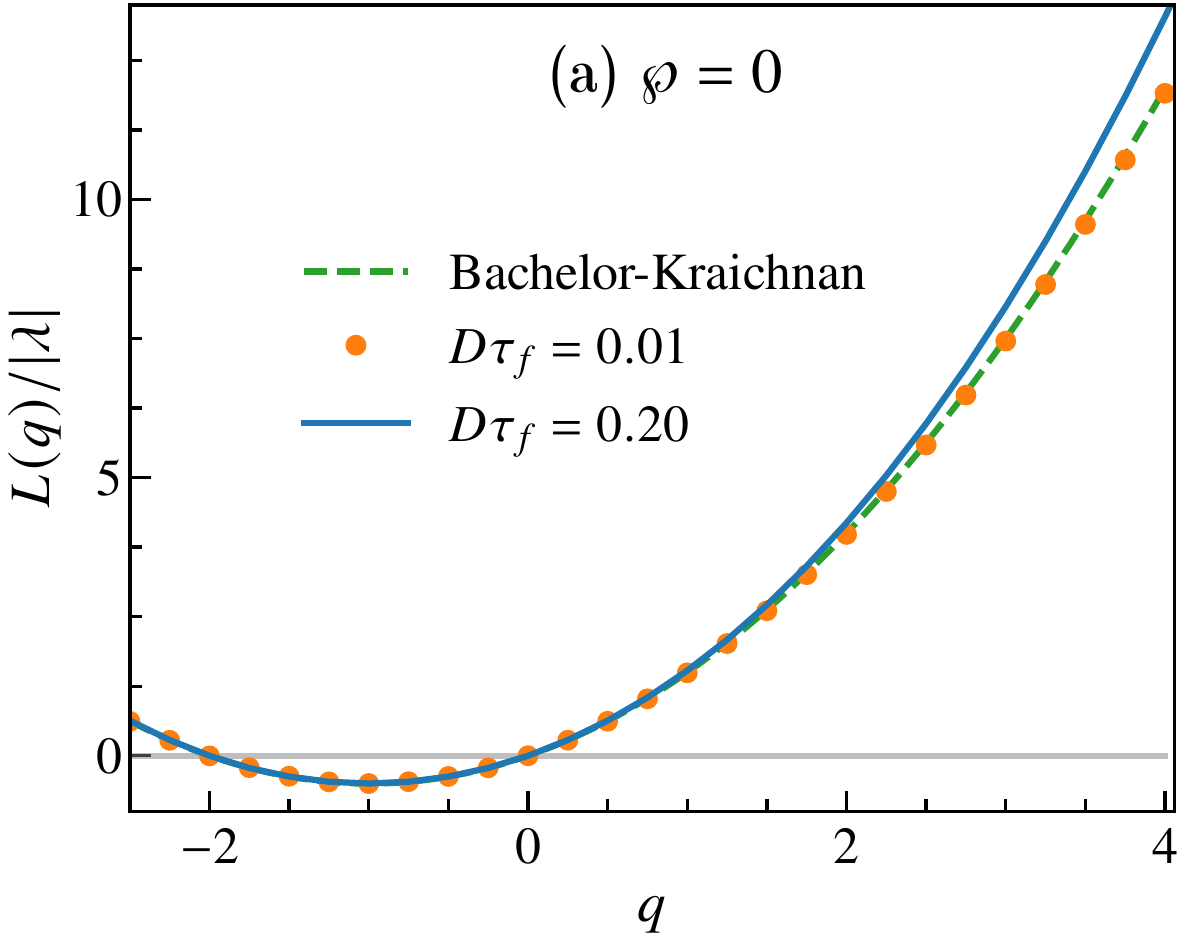}\hspace{2 em}
         \includegraphics[width=0.42\textwidth]{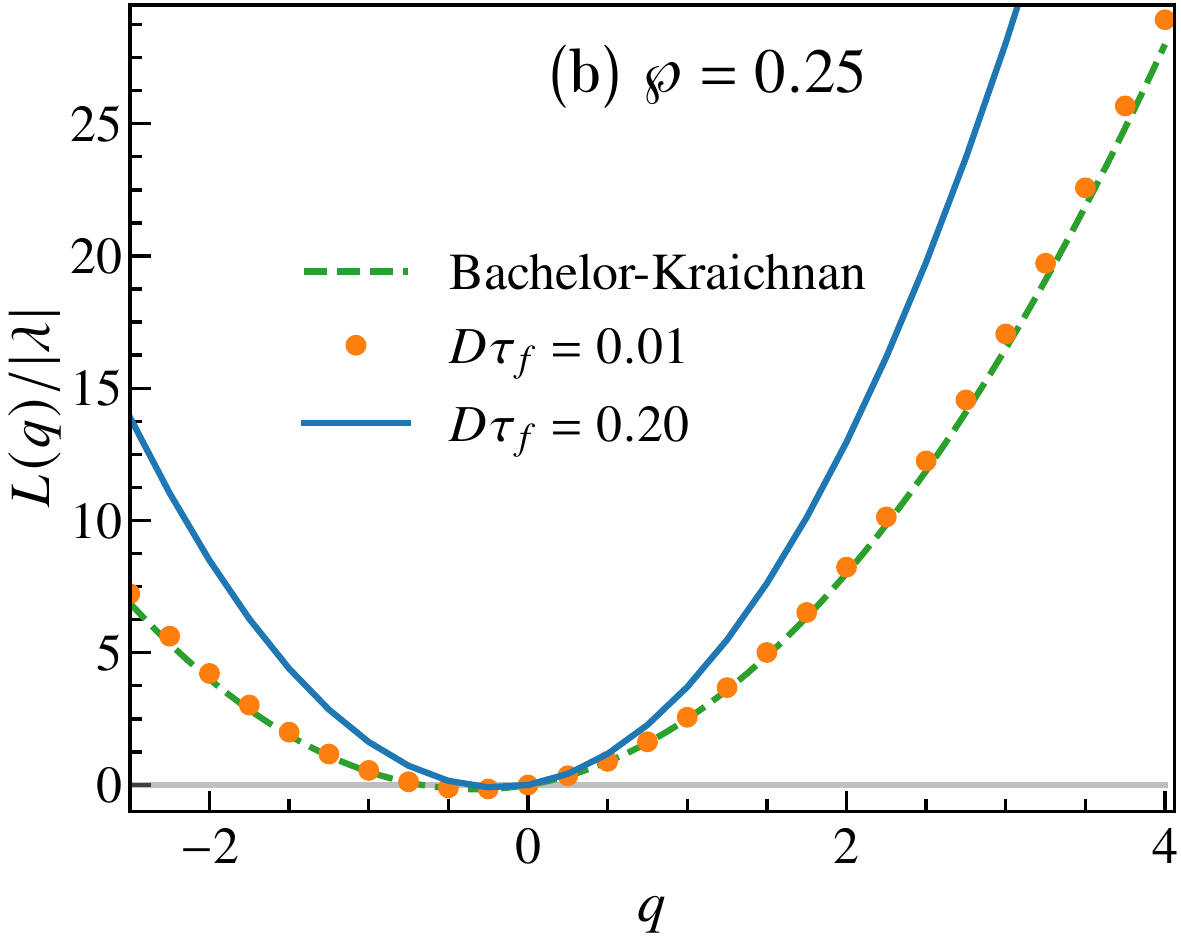}\\
         \includegraphics[width=0.42\textwidth]{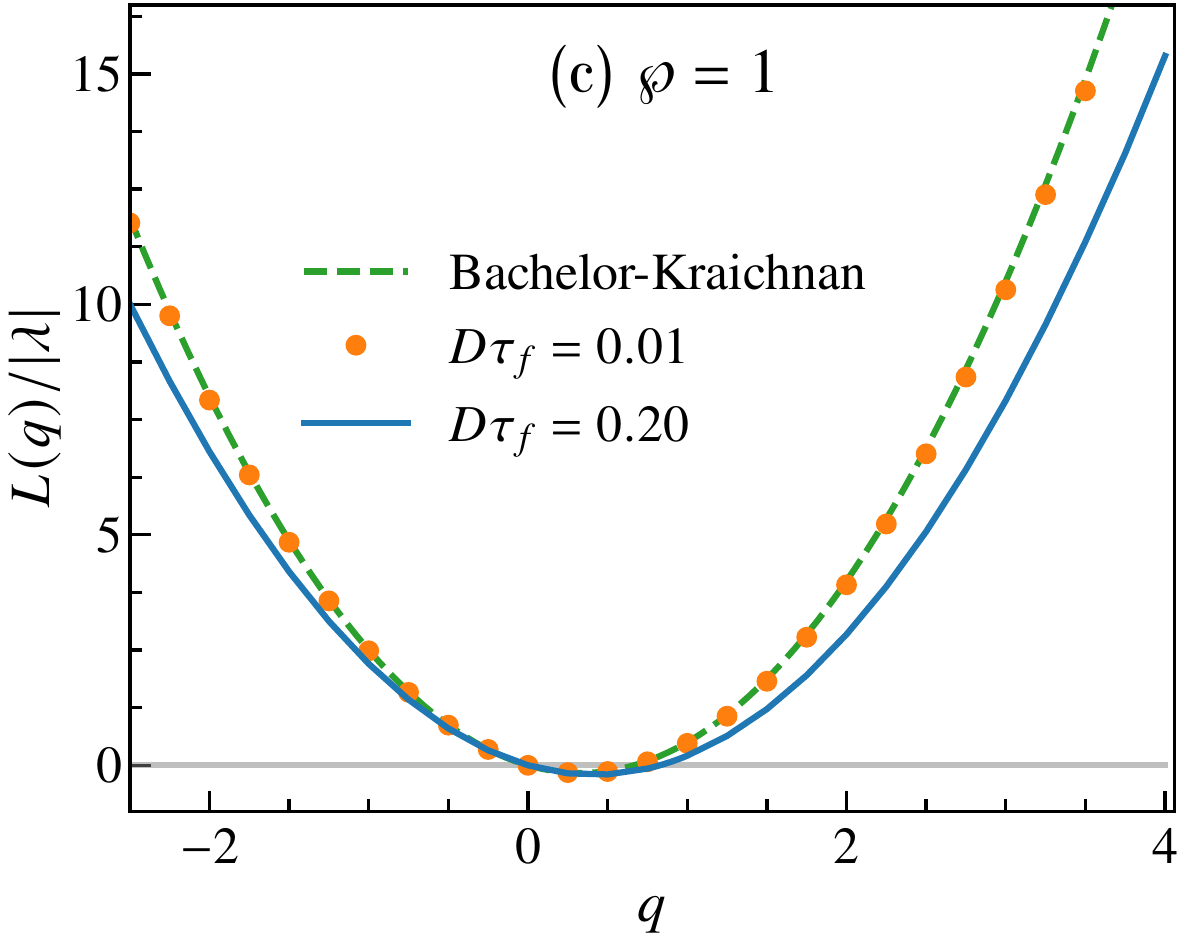}\hspace{2 em}  
         \includegraphics[width=0.42\textwidth]{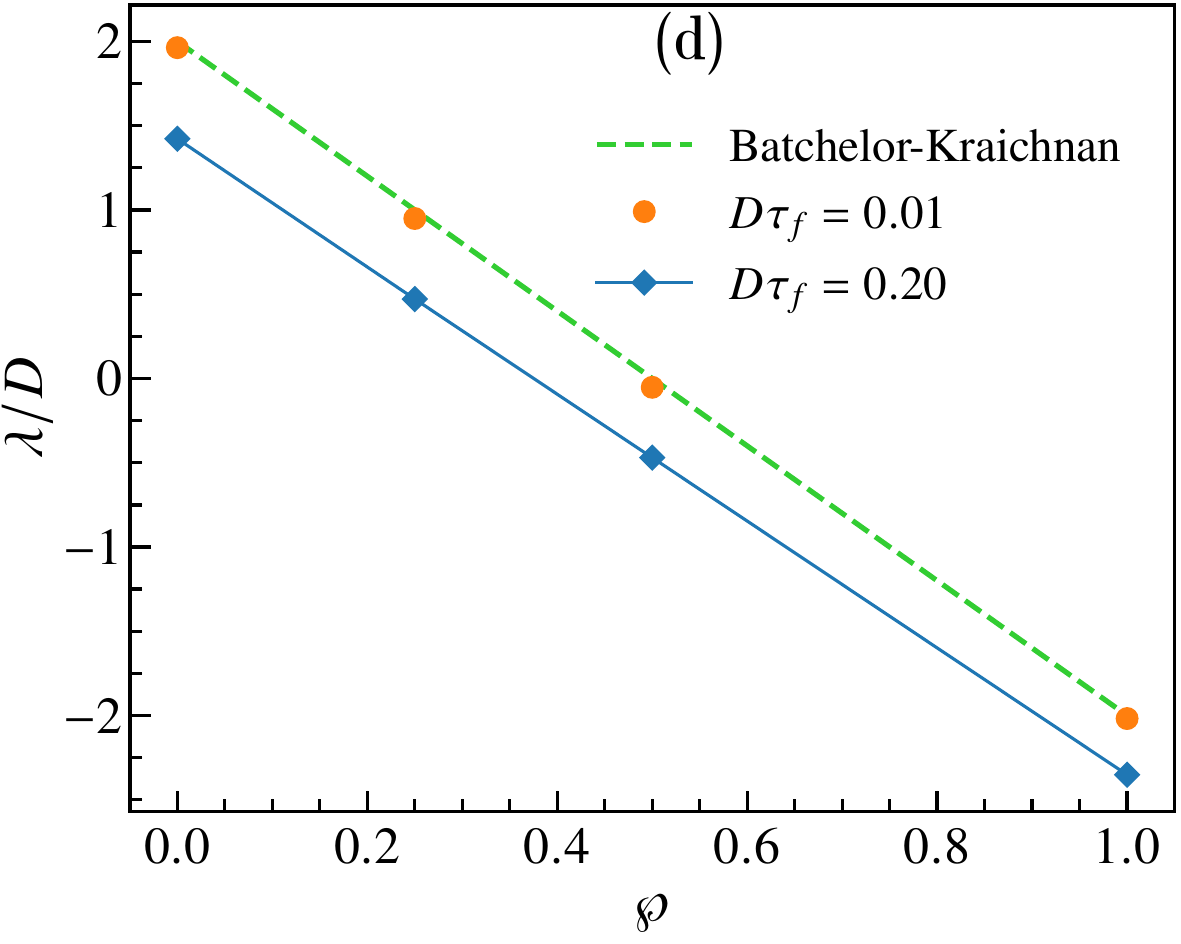}
         \caption{ $L(q)/\vert\lambda\vert$ vs $q$ for the two-dimensional Batchelor--Kraichnan model and the renewing flow with either $D\tau_f=0.01$ or $D\tau_f=0.2$. The degree of compressibility is ({\it a}) $\wp=0$, ({\it b}) $\wp=0.25$, and ({\it c}) $\wp=1$. Panel (d) presents the variation of $\lambda/D$ with $\wp$ in the two flows.
        }
        \label{fig:lqvsq}
    \end{figure}

We now check whether the effects of flow compressibility, identified in the previous section, remain the same in a flow with a finite time correlation. Specifically, we consider a two-dimensional renewing (or renovating) flow, where the velocity gradient along a Lagrangian trajectory is a sequence of independent identically-distributed matrices, each of which remains constant for a time $\tau_f$ \citep[see e.g.][]{zrms84,1995-childress-gilbert,y99}. The parameter $\tau_f$, thus, plays the role of the Lagrangian correlation time. We also account for the drift term of (\ref{eq:Stratonovich}), so that the results in the renewing flow approach those in the Batchelor--Kraichnan flow as  $\tau_f \to 0$. 
To be precise, the velocity gradient along a Lagrangian trajectory $\bm x(t)$ in the renewing flow is
\begin{equation}    \label{eq:defA}
\partial_j u_i(\bm x(t),t)=\mathsfi{G}_{ij}(t)-4\wp D\delta_{ij},
\end{equation}
where $D$ is a positive constant and
\begin{equation}
    \label{eq:G}
    \G(t)=\G_n,\;\;\mathrm{for}\;\; t\in\mathcal{I}_n=[t_n,t_{n+1}) \;\;\mathrm{with}\;\;t_n=n\tau_f, \; n\in\mathbb{N}.
\end{equation}
Here, $ \G_n=\ssf_n+\osf_n$ with 
\begin{equation}
    \label{eq:An}
    \ssf_n=
    \sqrt{\frac{2D}{\tau_f}}\,
    \begin{pmatrix}
        \sigma_{1,n} +\sqrt{2\wp}\,\sigma_{2,n} & \sigma_{3,n}
        \\
        \sigma_{3,n} & -\sigma_1 +\sqrt{2\wp}\,\sigma_{2,n}
    \end{pmatrix},
    \quad 
\osf_n=2\sqrt{\frac{D}{\tau_f}(1-\wp)}\,
    \begin{pmatrix}
        0 & \omega_n
        \\
        -\omega_n & 0
    \end{pmatrix}.
\end{equation}
In the above expressions, $\sigma_{i,n}$ ($i=1,2,3$ ) and $\omega_n$ are independent, zero-mean, Gaussian random variables such that
\begin{equation}
    \langle \sigma_{i,n}\sigma_{i,m}\rangle = 
    \langle \omega_{n}\omega_{m}\rangle = \delta_{n,m}.
\end{equation}
Hence, $\G(t)$ is Gaussian and zero-mean. Its two-time correlation is  
\begin{equation}
    \langle\mathsfi{G}_{ij}(t)\mathsfi{G}_{kl}(t')\rangle=2\mathsfi{C}_{ijkl}\, F(t,t')
\end{equation}
with
\begin{equation}
    F(t,t')=
    \begin{cases}
        \tau_f^{-1} & \text{if $\lfloor t/\tau_f\rfloor = \lfloor t'/\tau_f\rfloor$},
        \\
        0 & \text{otherwise},
    \end{cases}
\end{equation}
where $\lfloor \boldsymbol{\cdot}\rfloor$ denotes the floor function and $\mathsfi{C}_{ijkl}$ has been defined in \eqref{eq:C-tensor}. Thus, in the limit $D\tau_f\to 0$, the matrix $\G(t)$ has the same statistics as in the two-dimensional Batchelor--Kraichnan model  (note, indeed, that
$\int_{-\infty}^{\infty} F(t,t')\,dt'=1$). 
The form of $\G_n$ is obviously not unique, and other choices that reduce to the Batchelor--Kraichnan flow are also possible; however, the choice in \eqref{eq:An} has a minimal number of noises and is therefore advantageous in numerical simulations. 
Finally, the additional term $-4\wp D\mathsfbi{I}$ in \eqref{eq:defA} ensures that \eqref{eq:dtl} becomes equivalent to the It\^o equation~\eqref{eq:lG} as $D\tau_f\to 0$. Hence, in the limit $D\tau_f\to 0$, the dynamics of a line element in the renewing flow tends to that in the Batchelor--Kraichnan flow with same coefficient $D$ and same degree of compressibility $\wp$.
This convergence is illustrated in figure~\ref{fig:lqvsq} via a comparison of $L(q)$ (the calculation of $L(q)$ in the renewing flow is discussed below).

Our first goal is to determine $\alpha$, the power-law exponent of $\pst(R)$, and to verify the dynamical-systems theory of \S~\ref{sec:st-dist}. To this end, we compare the values of $\alpha$ obtained from simulations of the stochastic differential equation for $\bm R(t)$ to those calculated from \eqref{eq:alpha_l}. 
This comparison requires, first, the computation of $L(q)$, which is eased by the temporal properties of the renewing flow.
Since the velocity gradient 
in each interval $\mathcal{I}_n$ remains fixed, 
the solution of \eqref{eq:dtl} is
\begin{equation}
    \boldsymbol{\ell}(t_{n+1}) = \W_n\boldsymbol{\ell}(t_n),
\end{equation}
where
\begin{equation}
    \W_n
    =\mathrm{e}^{\tau_f(\G_n-4\wp D\mathsfbi{I})} = \mathrm{e}^{\sigma_{2,n}\sqrt{4D\tau_f\wp}-4D\tau_f\wp}\,\mathsfbi{M}_n
\end{equation}
with
\begin{equation}
\mathsfbi{M}_n=\begin{pmatrix}
        c_n+\textstyle\sqrt{\frac{2D\tau_f}{\Sigma_n}}\,\sigma_{1,n} s_n 
        &
        \sqrt{\frac{2D\tau_f}{\Sigma_n}}(\sigma_{3,n}+\sqrt{2}\omega_n\sqrt{1-\wp}) s_n 
        \\[2mm]
        \sqrt{\frac{2D\tau_f}{\Sigma_n}}(\sigma_{3,n}-\sqrt{2}\omega_n\sqrt{1-\wp}) s_n 
        &
        c_n-\textstyle\sqrt{\frac{2D\tau_f}{\Sigma_n}}\,\sigma_{1,n} s_n
    \end{pmatrix}.
\end{equation}
Here, $\Sigma_n = 2D\tau_f[\sigma_{1,n}^2+\sigma_{3,n}^2-2(1-\wp)\omega^2_n]$, and $c_n=\cosh(\sqrt{\Sigma_n})$, $s_n=\sinh(\sqrt{\Sigma_n})$ if  $\Sigma_n\geq0$ or $c_n=\cos(\sqrt{\vert\Sigma_n\vert})$,  $s_n=\sin(\sqrt{\vert\Sigma_n\vert})$ if $\Sigma_n<0$. 
The statistical isotropy of $\G(t)$ leads to the following formula for the Lyapunov and generalized Lyapunov exponents:
\begin{equation}
\label{eq:Lq-renewing}
    \lambda = \tau_f^{-1}\langle\ln\vert\W_n \bm e\vert^2\rangle,
    \quad
    L(q) = \tau_f^{-1}\ln\langle\vert\W_n\bm e\vert^{2q}\rangle,
\end{equation}
where $\bm e$ is any constant unit vector and the average is taken over the statistics of $\G_n$ (see e.g. \citealt{1995-childress-gilbert}, Ch.~11, and \citealt{y99}).
The averages in \eqref{eq:Lq-renewing} are calculated using the resampled Monte-Carlo method of \citet{2010-vanneste}; this method accelerates statistical convergence which would otherwise be extremely slow in the compressible case. (For large values of $\vert q\vert$ and values of $\wp$ close to unity, statistical convergence required up to $N=10^3$ time iterations and $K=16\times 10^6$ realizations of the renewing flow.)

The possibility of calculating $L(q)$ via \eqref{eq:Lq-renewing} is a significant simplification that is specific to renewing flows. 
For more complex chaotic flows, $L(q)$ must generally be calculated from direct numerical simulations of the Lagrangian dynamics; this can be an extremely challenging computation, especially for large $\vert q\vert$, because it requires accurate statistics of rare stretching events. Accurate knowledge of $L(q)$ for large positive $q$ is important while applying \eqref{eq:alpha_l} to stiff particles; we have found it necessary to go up to $q=4$, at least, in order to solve \eqref{eq:alpha_l} in the small-$D\tau_p$ regime.

    \begin{figure}
    \centering
\includegraphics[width=0.42\columnwidth]{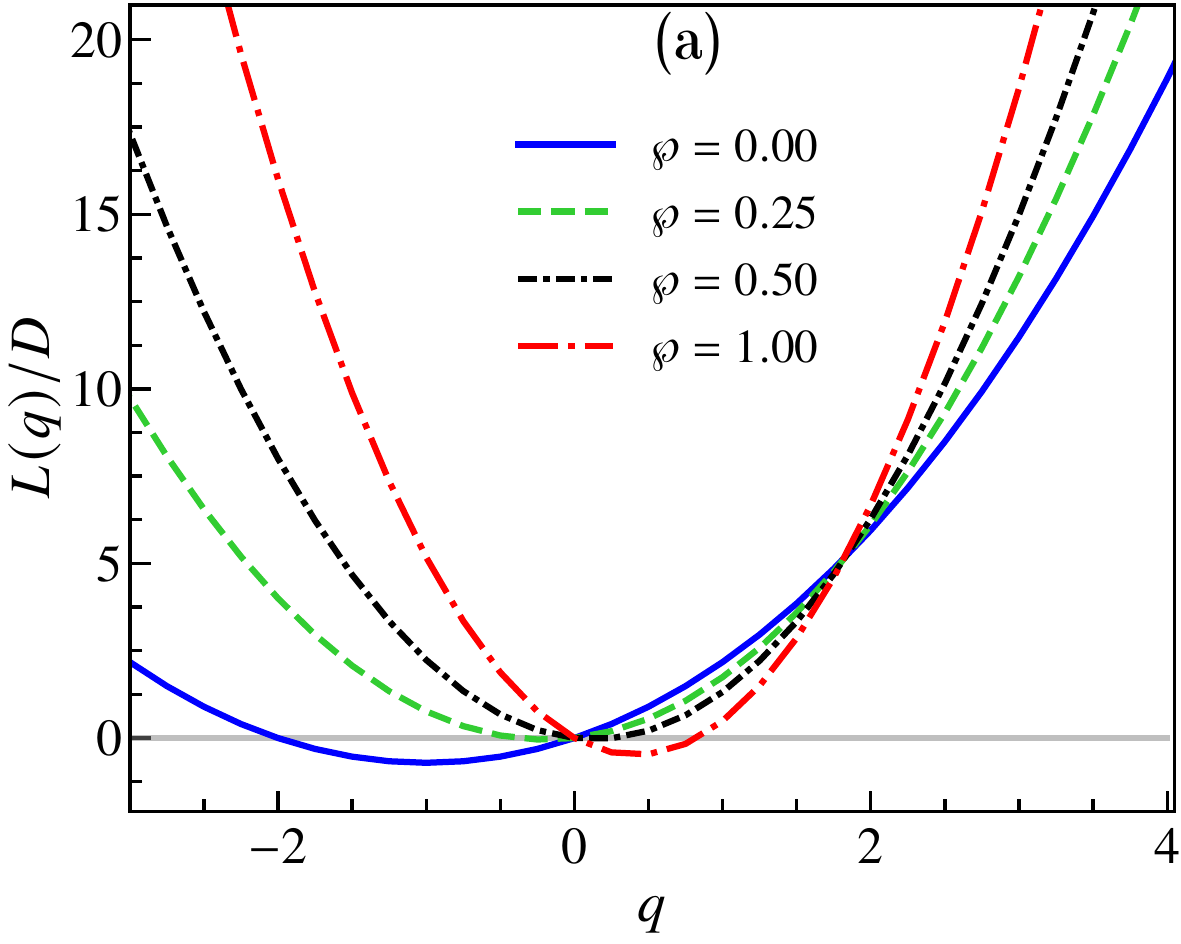}\hspace{2 em}%
\includegraphics[width=0.42\columnwidth]{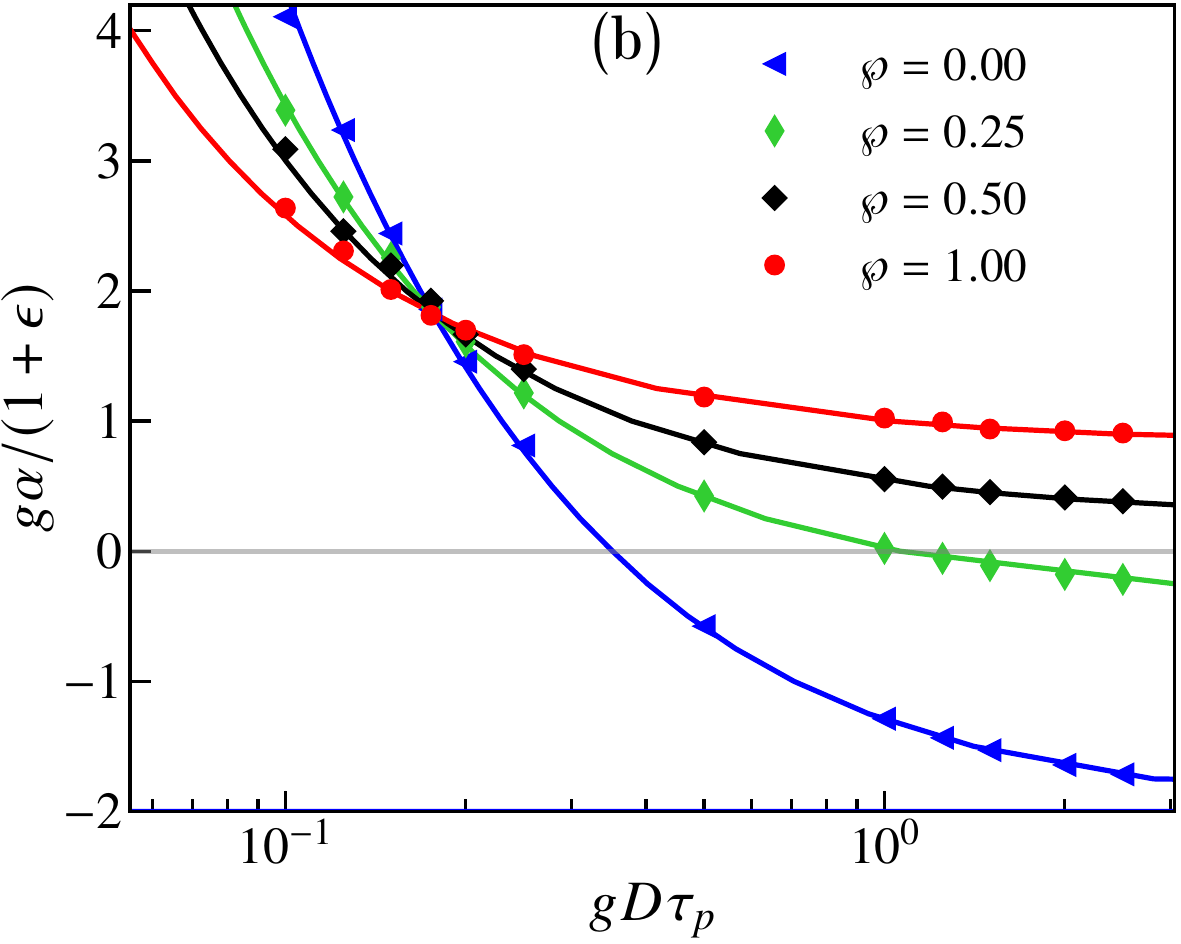}
        \caption{ Plots of ({\it a}) $L(q)/D$ as a function of $q$ and ({\it b}) $g \alpha /(1+\epsilon)$ vs $gD\tau_p$ for different values of $\wp$ in the renewing flow with $\tau_f=0.2$.    In panel ({\it b}), the lines correspond to the numerical solution of \eqref{eq:alpha_l}, whereas the symbols are estimates of the slope of $\pst(R)$ for $\Req\ll R\ll\Rmax$ obtained from numerical solutions of the stochastic differential equation for $\bm R(t)$.
        To obtain a nonzero equilibrium length and a finite maximum length, we modified \eqref{eq:orl} by adding Brownian noise with an amplitude such that $\Req=1$ and by replacing the linear elastic force with a nonlinear force that enforces $\Rmax=10^3$. For simplicity, we also took $\epsilon=0$ and $g=1$. The resulting stochastic differential equation for $\bm R(t)$ is given in \eqref{eq:dumbbell} and  \eqref{eq:fene}. We integrated \eqref{eq:dumbbell}  numerically by using the Euler-Maruyama method
        with time step $dt=5\times 10^{-4}$ in combination with \"Ottinger's rejection algorithm, to preserve the constraint $R<\Rmax$  (see \citealt{o96}, \S~4.3.2). These simulations confirm the validity of \eqref{eq:alpha_l} in the compressible case.}
        \label{fig:alpha_vs_wi} 
\end{figure}

In subsequent calculations, the correlation time is set to $\tau_f=0.2$ (unless stated otherwise), which for $\wp=0$ yields a Kubo number of $\mathit{Ku}\approx 0.3$. (For comparison, $\mathit{Ku}\approx 0.6$ in three-dimensional incompressible homogeneous isotropic turbulence; see \citealt{wg10}.)

Let us first examine the generalized Lyapunov exponents.
Figure~\ref{fig:lqvsq} shows that the stretching-rate statistics in the renewing flow, for $\tau_f=0.2$, deviates significantly from that in the Batchelor--Kraichnan flow, when both flows are compressible. ($L(q)$ remains quadratic, however, at least for the range of $q$ computed here.)
We find that
the critical degree of compressibility of the renewing flow, for which $\lambda=0$, is $\wp_c\approx 0.38$.
The variation of $L(q)$ with  $\wp$ is depicted in figure~\ref{fig:alpha_vs_wi}({\it a}); interestingly, $L(q)$ increases with $\wp$ for $q\gtrsim 2$, just as it does in the Batchelor--Kraichnan flow (see figure~\ref{fig:bk-plots}{\it a}).
So, even in the renewing flow, high-order moments of $\ell$ grow faster in time as the flow becomes more compressible, i.e.~large-stretching events become more prominent with increasing compressibility.
Consequently, the dependence of the exponent $\alpha$ on $D\tau_p$ and $\wp$ is analogous to that in the Batchelor--Kraichnan model (figure~\ref{fig:alpha_vs_wi}{\it b}).
In particular, compressibility has contrasting effects at small and large $gD\tau_p$: the exponent $\alpha$ decreases with $\wp$ at small $gD\tau_p$, while it increases with $\wp$ at large $gD\tau_p$;
the transition between the two behaviours occurs at $gD\tau_p\approx 0.18$.
The consequences of such variations of $\alpha$ for the stationary statistics of the particle size have already been discussed in \S~\ref{sec:BK}. 
Figure \ref{fig:alpha_vs_wi}({\it b}) also shows excellent agreement between the simulations and the theory of \S~\ref{sec:st-dist}.

\begin{figure}
\centering
\includegraphics[width=0.42\columnwidth]{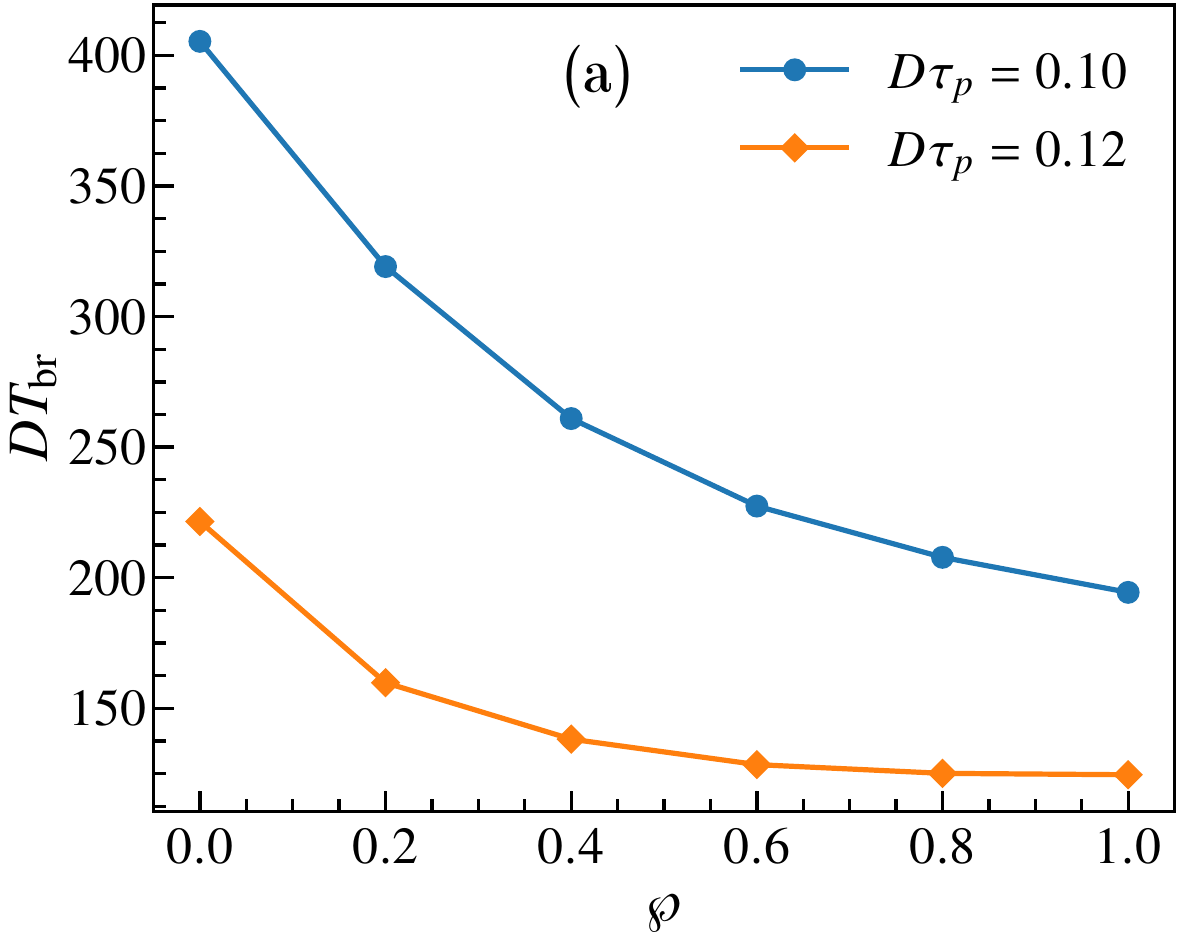}\\%
\includegraphics[width=0.42\columnwidth]{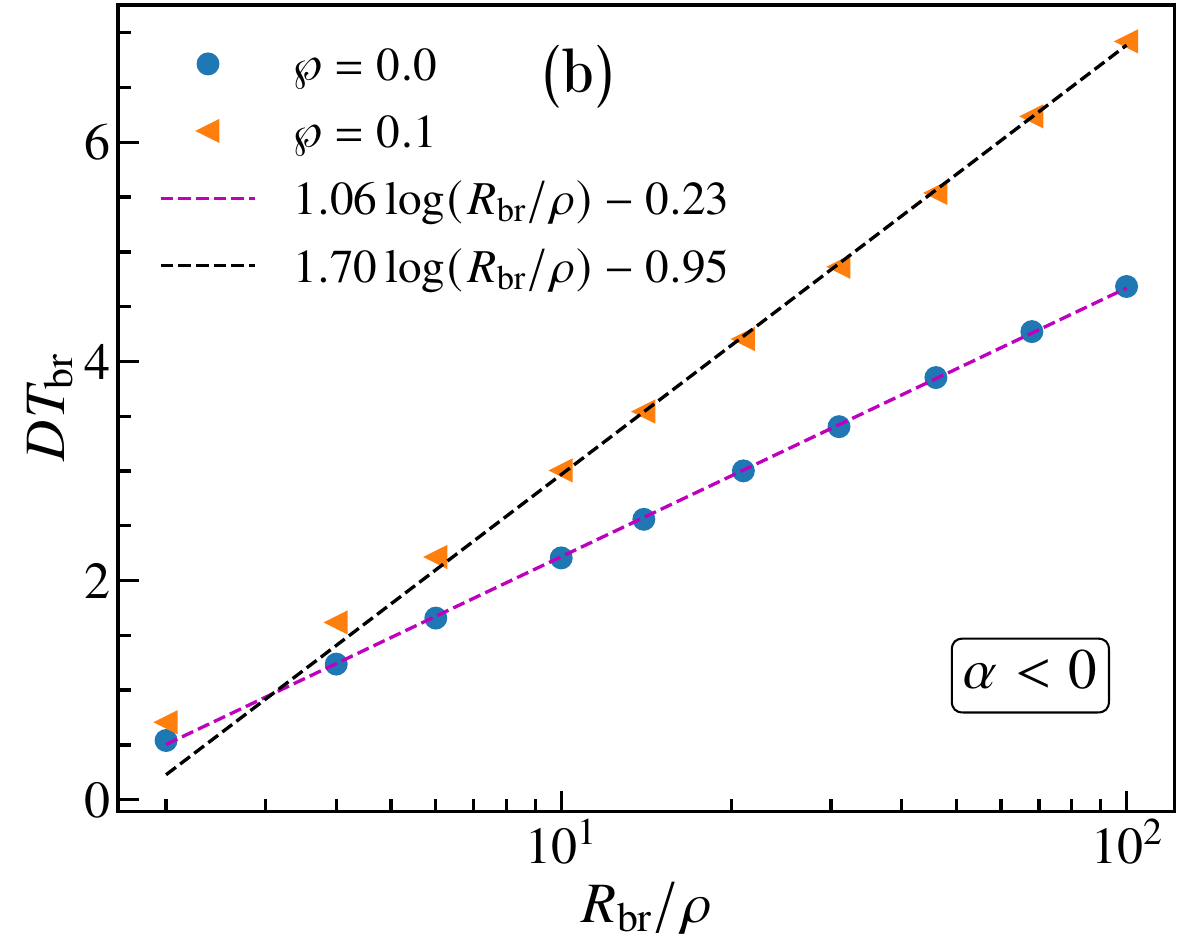}\hspace{2 em}%
\includegraphics[width=0.42\columnwidth]{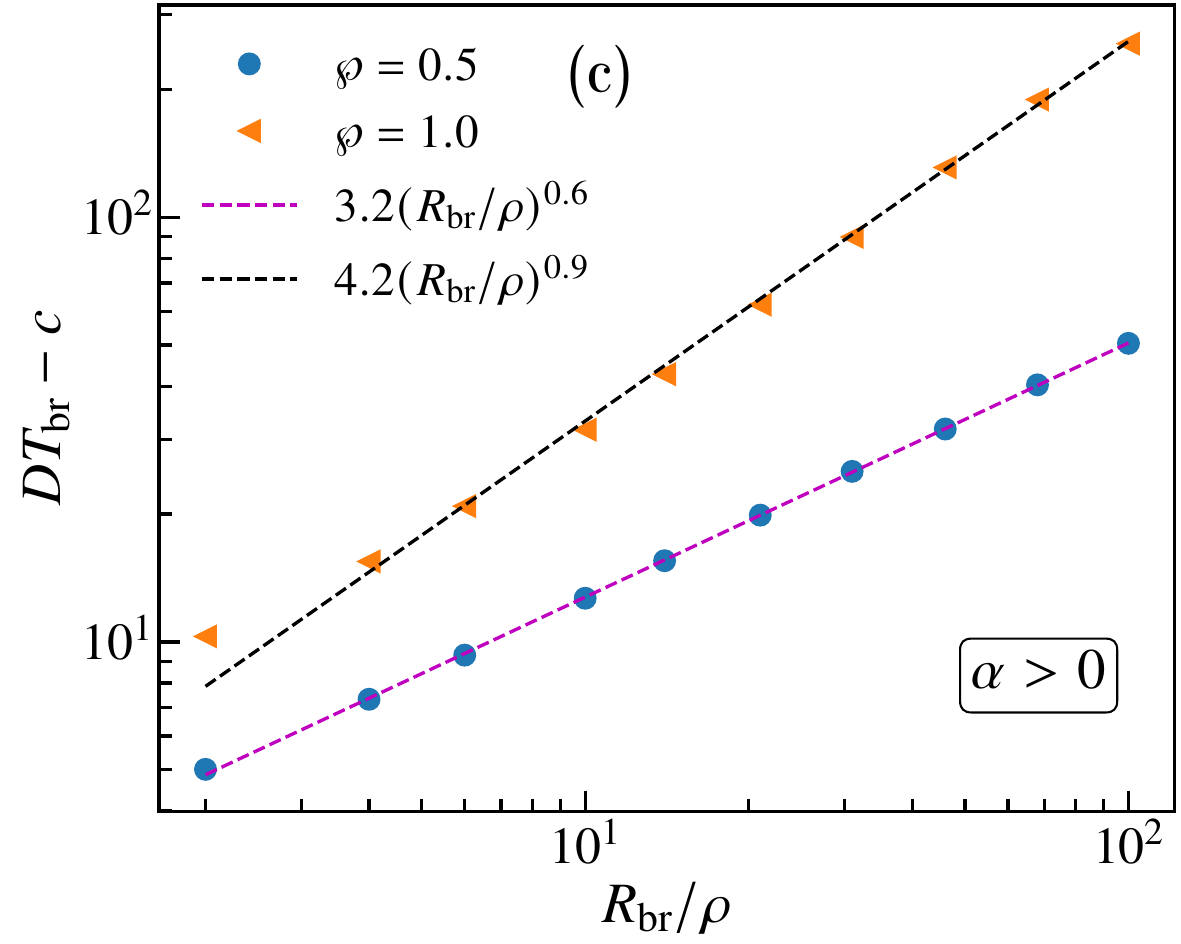}
        \caption{({\it a}) Rescaled mean breakup time as a function of the degree of compressibility for $\Rbr/\rho = 10$ and two different values of $D\tau_p$; ({\it b}) and ({\it c}) rescaled mean breakup time vs $\Rbr/\rho$ for $D\tau_p=1$ and different values of $\wp$. In panel ({\it c}), the constant $c$ is determined by the power-law fit of the data ($c =-3.4$ for $\wp=0.5$ and $c =-7.2$ for $\wp=1.0$). 
        In all panels, $\epsilon=0$ and $g=1$.}
        \label{fig:Tbr_renewing}
    \end{figure}

Next, we measure the dependence of the mean breakup time on the critical size and the degree of compressibility, by simulating the stochastic differential equation for $\bm R(t)$ (see the caption of figure~\ref{fig:alpha_vs_wi} for details).
As in the Batchelor--Kraichnan model, opposing variations of $D\Tbr$ with $\wp$ appear on either side of $gD\tau_p\approx 1.8$. At small $gD\tau_p$ (stiff particles), the mean breakup time reduces as the compressibility of the flow increases (figure~\ref{fig:Tbr_renewing}{\it a}). As explained in \S~\ref{sec:BK}, this surprising outcome can be understood by noting that (i) the dynamics of stiff particles are more sensitive to rare but intense straining events rather than to persistent but mild strains and (ii) large fluctuations of the straining rate become more prominent as $\wp$ increases.
At large $gD\tau_p$ (highly elastic particles), $D\Tbr$ is an increasing function of $\wp$, since breakup events are essentially determined by the mean dynamics (see figures~\ref{fig:Tbr_renewing}{\it b,c}). 

Regarding the dependence of $D\Tbr$ on the critical size for breakup, we find a transition from a logarithmic to a power-law dependence as $\alpha$ changes from negative to positive values (figures~\ref{fig:Tbr_renewing}{\it b,c}). This result 
agrees with the prediction of the Batchelor--Kraichnan model in  \eqref{eq:Tbr_asymptotic}.
In particular, the exponents of the power laws in figure~\ref{fig:Tbr_renewing}({\it c}) are close to the values of $\alpha$ that determine the slope of $\pst(R)$ in the absence of breakups (figure~\ref{fig:alpha_vs_wi}{\it b}).

\section{Concluding remarks}
\label{sec:con}

We have examined the extensional dynamics of tiny deformable particles in compressible fluctuating flows, using a generic particle model and simple, synthetic, random flows. In these flows, the mean stretching of line elements is suppressed as compressibility increases, as evidenced by the decreasing value of the principal Lyapunov exponent $\lambda$. This mean behaviour, however, does not determine entirely the effect of compressibility on particle stretching and breaking. Indeed, a key insight of this work is that the fluctuations of the line-element stretching rate play a critical role, giving rise to several important effects: (i) a broad, power-law distribution of extension, which persists even when the flow is strongly compressible (with $\lambda < 0$); this implies that, regardless of the degree of compressibility, large deviations of the stretching rate can produce highly-extended particles; (ii) a counter-intuitive increase with compressibility of the stretching and breakup rate of stiff particles, provided the large positive fluctuations of the stretching rate---quantified by high positive-order generalized Lyapunov exponents---increase with compressibility; (iii) a reduction in the sharpness of the shrink--stretch transition;
this outcome follows from the aforementioned increase in the stretching of stiff particles, as well as a contrasting decrease in the stretching of highly-elastic particles.

The opposing effects of compressibility on stiff and highly-elastic particles are due to (i) the increase in the frequency of intense-straining events as the compressibility of our model flows is increased and (ii) the differing impact of such events on stiff and highly-elastic particles. The long elastic relaxation time of highly-elastic particles averages out the strain rate and renders the extension insensitive to short bursts of straining. In contrast, stiff particles with a short relaxation time are more responsive to sudden strong straining; in fact, such events are necessary to overcome the strong elastic restoring force and stretch out the particle; so an increase in the frequency of strong-straining events, even at the cost of suppressing the mean straining, will produce more extended particles.

The two model flows considered here, namely the time-decorrelated Batchelor--Kraichnan flow and the time-correlated renewing flow, both exhibit an increase with flow compressibility of the magnitude of the high positive-order generalized Lyapunov exponents. There is no reason why this should be true in general, and it is quite possible that other flows show the opposite trend; if so, then the frequency of strong-straining events will decrease with compressibility and both stiff and highly-elastic particles will stretch less in flows of increasing compressibility. This qualitative dependence of particle stretching on the full statistics of straining is the crucial insight of the present work.


Predicting the effect of compressibility on deformable particles in a given flow will require knowledge of the variation of the generalized Lyapunov exponents with compressibility. Since computing these exponents is, in general, very challenging, it would be expedient to directly simulate the dynamics of deformable particles in the flow of interest. Even so, the physical connection between the generalized Lyapunov exponents and particle stretching provides a rationale for understanding the behaviour of particles in different kinds of compressible flows.


We have limited our study to flows with just two values of the temporal correlation, corresponding to $\mathit{Ku} = 0$ (delta-correlated) and $\mathit{Ku} = 0.3$. 
Our goal here was just to check whether the analytical predictions made using the delta-correlated flow continued to hold for a flow with a finite correlation time. The quantitative details of the particle dynamics will vary with $\mathit{Ku}$ and an investigation of this dependence could be taken up in future work. (Our limited explorations suggest that the results will change significantly only for $\mathit{Ku} \gg 1$; this case, however, corresponds to a near steady flow and so is not of interest in the context of fluctuating flows.) A task that demands more immediate attention is the study of particle deformation in direct numerical simulations, say of a surface flow, to see how our qualitative predictions manifest in a turbulent suspension. 

\begin{bmhead}[Acknowledgments.]
J.R.P. acknowledges his Associateship with the International Centre for Theoretical Sciences (ICTS), Tata Institute of Fundamental Research, Bengaluru,
India. D.R.~and D.V.~are grateful to the Universit\'e C\^ote d’Azur’s Center for High-Performance Computing (OPAL infrastructure) for providing resources and support.
\end{bmhead}

\begin{bmhead}[Funding information.]
This work was supported by 
the Indo-French Centre for the Promotion of Advanced Research (IFCPAR/CEFIPRA, Project No. 6704-1)
and by the French government through the France 2030 investment plan managed by the National Research Agency (ANR), as part of the Initiative of Excellence Universit\'e C\^ote d’Azur under reference number ANR-15-IDEX-01.
D.V. acknowledges
the support of Agence Nationale de la Recherche through Project No.~ANR-21-CE30-0040-01. 
\end{bmhead}

\begin{bmhead}[Declaration of interests.]
The authors report no conflict of interest.
\end{bmhead}

\appendix

\section{PDF of the particle size in the compressible Batchelor--Kraichnan model}
\label{sec:app-1}

Here, we calculate the exact PDF of $R$, over the entire interval $[0,\Rmax]$, and see how the power-law behaviour (discussed in Section~\ref{sec:BK}) appears over intermediate extensions $\Req\ll R\ll \Rmax$. This requires \eqref{eq:orl} to be modified in order to regularize the behaviour at small and large extensions. We therefore introduce thermal noise, which sets the equilibrium size $\Req$, and enforce the maximum extension $\Rmax$, by making the elastic force nonlinear.  
Restricting ourselves to the case of $\epsilon=0$ and $g=1$, for simplicity, we obtain
\begin{equation}
\label{eq:dumbbell}
\dot{R}_i=R_j\partial_j u_i-\frac{f(R)R_i}{2\tau_p} + \sqrt{\dfrac{\Req^2}{\tau_p d}}\,\xi_i(t), 
\end{equation}
where $\bm\xi(t)$ is a $d$-dimensional white noise and 
\begin{equation} f(R)=\dfrac{\Rmax^2}{\Rmax^2-R^2}.
\label{eq:fene}
\end{equation}
Equation~\eqref{eq:dumbbell} corresponds to the finitely extensible nonlinear elastic (FENE) dumbbell model, well-studied in polymer physics \citep{bird}.
As discussed in \S~\ref{sec:BK},
\eqref{eq:dumbbell} is equivalent to the It\^o
stochastic differential equation
\begin{equation}
\label{eq:dumbbell-G}
\dot{\bm R}=\G(t)\bm R-\frac{f(R)\bm R}{2\tau_p} + \sqrt{\dfrac{\Req^2}{\tau_p d}}\,\bm \xi(t),  \end{equation}
where $\G(t)$ has been defined in \eqref{eq:corr-G}.
The PDF of $\bm R$ at time $t$, denoted as $\mathscr{P}(\bm R,t)$, satisfies the Fokker--Planck equation~\citep*{pav16}
\begin{subequations}
\label{eq:FPE_cartesian}
\begin{align}
\dfrac{\partial\mathscr{P}}{\partial t} &=
\mathsfi{C}_{ijkl}
\dfrac{\partial^2}{\partial R_i\partial R_k} (R_jR_l\mathscr{P})
+
\dfrac{1}{2\tau_p}\,\dfrac{\partial}{\partial R_i}[f(R)R_i\mathscr{P}]
+
\dfrac{\Req^2}{2\tau_p d}\,
\nabla^2_{\bm R}\mathscr{P}
\\
&=D(d+1-2\wp)\dfrac{\partial^2}{\partial R_i\partial R_i}(R^2\mathscr{P})+2D(\wp d-1)
\dfrac{\partial^2}{\partial R_i\partial R_j}(R_i R_j\mathscr{P})
\\ \nonumber
& \hspace{12em}+
\dfrac{1}{2\tau_p}\,\dfrac{\partial}{\partial R_i}[f(R)R_i\mathscr{P}]
+
\dfrac{\Req^2}{2\tau_p d}
\,\nabla^2_{\bm R}
 \mathscr{P},
\end{align}
\end{subequations}
where summation over repeated indices is implied.
Assuming that the initial condition of \eqref{eq:FPE_cartesian} is independent of the orientation of $\bm R$, we expect the solution of \eqref{eq:FPE_cartesian} to retain this property at all times, since the velocity gradient is statistically isotropic. To take advantage of this fact,
we transform from Cartesian variables $\bm R$ to azimuthal variables $(R,\Psi)$, where
$\Psi$ denotes a suitable set of angles in $d$ dimensions.
The PDF $\widetilde{\mathscr{P}}(R,\Psi,t)$ is connected to $\mathscr{P}(\bm R,t)$ via the relation $\widetilde{\mathscr{P}}(R,\Psi,t)=J \mathscr{P}(\bm R,t)$, where $J$ is the Jacobian of the transformation between the two sets of variables and is of the form 
$J=R^{d-1}h(\Psi)$ with 
$h(\Psi)$ being a function of the angles.  Owing to statistical isotropy, the dependence of 
$\widetilde{\mathscr{P}}(R,\Psi,t)$ on the angles is contained entirely in $J$; therefore,
$\widetilde{\mathscr{P}}(R,\Psi,t)=P(R,t)h(\Psi)$, 
where $P(R,t)$ is the marginal PDF of $R$.
This implies that
$\mathscr{P}(\bm R,t)=R^{1-d}P(R,t)$.
The
Fokker--Planck equation for 
$P(R,t$) is obtained
from \eqref{eq:FPE_cartesian} 
by changing the variables in the partial derivatives, 
discarding the derivatives with respect to the angular variables, and
replacing $\mathscr{P}$ with $R^{1-d}P$:
\begin{equation}\label{eq:FP}
    \partial_t P = - \partial_R  [\mathfrak{D}_1(R) P ] + \partial_R^2 [\mathfrak{D}_2(R) P]
\end{equation}
where
\begin{equation}
\label{eq:D1_D2_BK}  
 \mathfrak{D}_1( R)=\left(\lambda+\dfrac{\Delta}{2}\right)R
-\dfrac{f(R)R}{2\tau_p}+\dfrac{(d-1)\Req^2}{2\tau_p d\, R}, 
\qquad
\mathfrak{D}_2( R)=\dfrac{\Delta}{2}\, R^2
 +\dfrac{\Req^2}{2\tau_p d}.
\end{equation}
%
With reflecting boundary conditions at both ends of the interval $0\leqslant R \leqslant \Rmax$ (which set the probability current $\partial_{R} (\mathfrak{D}_2 P) - \mathfrak{D}_1 P$ to zero at $R=0$ and $R=\Rmax$), \eqref{eq:FP} has the stationary solution
$\pst(R) = e^{-\Phi_a(R)}$,
where $\Phi_a(R)$ has been defined in \eqref{eq:Phi}. Using the expressions of the drift and diffusion  coefficients in \eqref{eq:D1_D2_BK} yields
\begin{equation} \pst(R) \propto R^{d-1}
\left(1+ d\Delta\tau_p\dfrac{R^2}{\Req^2}\right)^{-\frac{\alpha+d}{2}}\left(1-\frac{R^2}{\Rmax^2}\right)^{\frac{\alpha}{2}+\frac{\lambda}{\Delta}}
\end{equation}
with
\begin{equation}
\alpha = -\dfrac{2\lambda}{\Delta}+\dfrac{bd}{1+bd\Delta\tau_p}, \qquad
b=\dfrac{\Rmax^2}{\Req^2}.
\end{equation}
In the range $\Req\ll R\ll\Rmax$, 
the stationary PDF behaves as $R^{-1-\alpha}$ and,
in the limit $b\to\infty$, the exponent $\alpha$ becomes
\begin{equation}
\alpha=\dfrac{1}{\Delta}\left(\dfrac{1}{\tau_p}-2\lambda\right).
\end{equation}

\section{Zero $\tau_f$ limit of the renewing flow}

Consider the equation 
\begin{equation}
\dfrac{\mathrm{d}\bm \ell}{\mathrm{d}t}=\G(t)\bm\ell
\end{equation}
with $\G(t)$ as in \eqref{eq:G}. In each time interval $\mathcal{I}_n$, its solution is 
\begin{equation}
\bm\ell(t_{n+1})=\mathrm{e}^{\tau_f\G_n}\bm\ell(t_n).
\end{equation}
Hence 
\begin{equation}
\bm\ell(t_{n+1})=\mathrm{e}^{\tau_f\G_n}\mathrm{e}^{\tau_f\G_{n-1}} \cdots\mathrm{e}^{\tau_f\G_2} \mathrm{e}^{\tau_f\G_1} \bm\ell(0).
\end{equation}
As $D\tau_f\to 0$, $\G(t)$ tends to the $d\times d$ white-noise defined in \eqref{eq:corr-G}, while $\bm\ell(t)$ tends to the time ordered exponential  
\begin{equation}
\bm\ell(t)=\mathcal{T}\exp\left(\int \G(t)\, \mathrm{d}t\right),
\end{equation}
which is the solution of the Stratonovich equation 
\begin{equation}
\dfrac{\mathrm{d}\bm\ell}{\mathrm{d}t}=\G(t)\circ\bm\ell.
\end{equation}
To recover the Stratonovich equation equivalent to \eqref{eq:lG} in the limit $D\tau_f\to 0$, it is therefore necessary to subtract the drift term $4\wp D\mathsfbi{I} \bm\ell$ 
from the right-hand side of the evolution equation for $\bm\ell(t)$ (see \eqref{eq:Stratonovich}).


\bibliographystyle{jfm-FLM-arxiv}
\bibliography{polymers}

@book{gardiner,
  author    = {C. W. Gardiner}, 
  title     = {Handbook of Stochastic Methods},
  publisher = {Springer},
  year      = 1985,
  address   = {Berlin, Heidelberg},
}

@article{coletti25,
  title = {Spatiotemporal scales of motion and particle clustering in free-surface turbulence},
  author = {Li, Y. and Sanness Salmon, H. and Hassaini, R. and Chang, K. and Mucignat, C. and Coletti, F.},
  journal = {Phys. Rev. Fluids},
  volume = {10},
  issue = {3},
  pages = {034602},
  numpages = {19},
  year = {2025},
  month = {Mar},
  publisher = {American Physical Society},
  doi = {10.1103/PhysRevFluids.10.034602},
  url = {https://link.aps.org/doi/10.1103/PhysRevFluids.10.034602}
}

@book{1995-childress-gilbert,
    title       =   {Stretch, Twist, Fold: the Fast Dynamo},
    series      =   {Lecture Notes in Physics Monographs},
    author      =   {Childress, S. and Gilbert, A. D.},
    volume      =   {37},
    year        =   {1995},
    publisher   =   {Springer Science \& Business Media},
    doi         =   {https://doi.org/10.1007/978-3-540-44778-8}
}

@article{2010-vanneste,
    title       = {Estimating generalized {L}yapunov exponents for products of random matrices},
    author      = {Vanneste, J.},
    journal     = {Phys. Rev. {\rm E}},
    volume      = {81},
    pages       = {036701},
    numpages    = {12},
    year        = {2010},
    publisher   = {American Physical Society},
    doi         = {10.1103/PhysRevE.81.036701}
}

@article{2018-ray-vincenzi, 
    title       =   {Droplets in isotropic turbulence: deformation and breakup statistics}, 
    volume      =   {852}, 
    DOI         =   {10.1017/jfm.2018.453}, 
    journal     =   {J. Fluid Mech.}, 
    author      =   {Ray, S. S. and Vincenzi, D.}, 
    year        =   {2018}, 
    pages       =   {313}
}

@article{2023-cunha-etal, 
    title       =   {Flow classification from the perspective of microelements dispersed in a continuous phase}, 
    volume      =   {321}, 
    journal     =   {J. Non-Newtonian Fluid Mech.}, 
    author      =   {J. P. Cunha and P. R. de Souza Mendes and R. L. Thompson and E. C. Rodrigues and E. F. Quintella}, 
    year        =   {2023}, 
    pages       =   {105094},
    doi         =   {http://dx.doi.org/10.2139/ssrn.4379567}
}

@article{ppv23,
author = {J.~R.~Picardo and E.~.L.~C. VI M.~Plan and D.~Vincenzi},
title = {Polymers in turbulence: stretching statistics and the role of extreme strain rate fluctuations},
year = {2023},
journal = {J. Fluid Mech.},
volume = {969},
pages = {A24},
doi = {https://doi.org/10.1017/jfm.2023.524}
}

@article{tanner76,
author = {R. J. Tanner},
title = {A test particle approach to flow classification for viscoelastic fluids},
year = {1976},
journal = {AIChE J.},
volume = {22},
pages = {910--918},
doi = { https://doi.org/10.1002/aic.690220515}
}

@inproceedings{y99,
author = {W. R. Young}, 
title =  {Stirring and Mixing},
booktitle = {1999 Summer Program in Geophysical Fluid Dynamics},
year = {1999},
editor = {J.-L. Thiffeault and C. Pasquero}, 
organization = {Woods Hole Oceanographic Institution},
address = {Woods Hole, MA}
}

@incollection{g08,
	author = "K. Gaw\c{e}dzki",
	title = "Soluble models of turbulent transport",
	booktitle = "Non-equilibrium Statistical Mechanics and Turbulence",
	year = "2008",
	editor = "S. Nazarenko and  O. V. Zaboronski",
	pages = "44--107",
	publisher = "Cambridge University Press",
	series = "London Mathematical Society Lecture Note Series",
	volume = "355",
        doi = {https://doi.org/10.1017/CBO9780511812149}
}

@article{sl94,
journal = "Chaos Solitons Fract.",
volume = "4",
pages = "913",
year = "1994",
author = "A. J. Szeri and L. G. Leal",
title = "Orientation Dynamics and Stretching of Particles in Unsteady, Three-dimensional Fluid Flows: Unsteady Attractors",
doi = {https://doi.org/10.1016/0960-0779(94)90131-7}
}

@article{swl91,
journal = "J. Fluid Mech.",
volume = "228",
pages = "207--241",
year = "1991",
author = "A. J. Szeri and S. Wiggins and L. G. Leal",
title = "On the dynamics of suspended microstructure in unsteady, spatially inhomogeneous, two-dimensional fluid flows",
doi = {https://doi.org/10.1016/0960-0779(94)90131-7}
}

@article{ko86,
journal = "J. Non-Newtonian Fluid Mech.",
volume = "21",
pages = "127--131",
year = "1986",
author = "D. V. Khakhar and  J. M Ottino",
title = "A note on the linear vector model of {Olbricht, Rallison, and Leal} as applied to the breakup of slender axisymmetric drops",
doi = "https://doi.org/10.1016/0377-0257(86)80067-2"
}

@article{sl92,
journal = "J. Fluid Mech.",
volume = "242",
pages = "549--576",
year = "1992",
author = "A. J. Szeri and L. G. Leal",
title = "A new computational method for the solution of flow problems of microstructured fluids. {P}art 1. {T}heory",
doi = {10.1017/S0022112092002490}
}

@article{sl93,
journal = "J. Fluid Mech.",
volume = "250",
pages = "143--167",
year = "1993",
author = "A. J. Szeri and L. G. Leal",
title = "Microstructure suspended in three-dimensional flows", 
doi={10.1017/S0022112093001417}
}

@article{va97,
journal = "Physica {\rm D}",
volume = "106",
pages = "148",
year = "1997",
author = "M. Vergassola and M. Avellaneda",
title = "Scalar transport in compressible flow",
doi = {https://doi.org/10.1016/S0167-2789(97)00022-5}
}

@article{bdes04,
title = "Lagrangian tracers on a surface flow: the role of time correlations",
journal = "Phys. Rev. Lett.",
volume = "93",
pages = "134501",
year = "2004",
author = "G. Boffetta and J. Davoudi and B. Eckhardt and J. Schumacher",
doi = {https://doi.org/10.1103/PhysRevLett.93.134501}
}

@article{coletti23,
title = "Effect of shape and size on the transport of floating particles on the free surface in a natural stream",
journal = "Water Resour. Res.",
volume = "59",
pages = "e2023WR035716",
year = "2023",
author = "H. R. {Sanness Salmon} and L. J. Baker and J. L. Kozarek and F. Coletti",
doi = {https://doi.org/10.1029/2023WR035716}
}

@article{cdgs04,
title = "Eulerian and {Lagrangian} studies in surface flow turbulence",
journal = "New J. Phys.",
volume = "6",
pages = "53",
year = "2004",
author = "J. R. Cressman and J. Davoudi and W. I. Goldburg and J. Schumacher",
doi = {https://dx.doi.org/10.1088/1367-2630/6/1/053},
}

@article{so93,
journal = "Science",
volume = "259",
pages = "335--339",
year = "1993",
author = "J. C. Sommerer and E. Ott",
title = "Particles floating on a moving fluid: A dynamically comprehensible physical fractal",
URL = {http://www.jstor.org/stable/2880917}
}

@article{bmv14,
journal = "J. Fluid Mech.",
volume = "754 ",
pages = "184",
year = "2014",
author = "L. Biferale and C. Meneveau and R. Verzicco",
title = "Deformation statistics of sub-{K}olmogorov-scale ellipsoidal neutrally buoyant drops in isotropic turbulence",
doi = {https://doi.org/10.1017/jfm.2014.366}
}

@article{bdl06,
title = "Multifractal clustering of passive tracers on a surface flow",
journal = "Europhys. Lett.",
volume = "74",
pages = "62--68",
year = "2006",
author = "G. Boffetta and J. Davoudi and F. {De Lillo}",
doi = {https://dx.doi.org/10.1209/epl/i2005-10503-6}
}

@article{p-m14,
title = "Mixing and clustering in compressible chaotic stirred flows",
journal = "Phys. Rev. {\rm E}",
volume = "89",
pages = "022917",
year = "2014",
author = "Vicente {P\'erez--Mu\~{n}uzuri}",
doi = {https://doi.org/10.1103/PhysRevE.89.022917}
}

@article{jhbm17,
journal = "Phys. Rev. Fluids",
volume = "2",
pages = "014605",
year = "2017",
title = "Analysis of geometrical and statistical features of {Lagrangian} stretching in turbulent channel flow using a database task-parallel particle tracking algorithm",
author = "P. L. Johnson and S. S. Hamilton and R. Burns and C. Meneveau",
doi = {https://doi.org/10.1103/PhysRevFluids.2.014605}
}

@article{jm16,
journal = "Phys. Rev. {\rm E}",
volume = "93",
pages = "033118",
year = "2016",
title = "Large-deviation statistics of vorticity stretching in isotropic turbulence",
author = "P. L. Johnson and C. Meneveau",
doi = {https://doi.org/10.1103/PhysRevE.93.033118}
}

@article{jm15,
journal = "Phys. Fluids",
volume = "27",
pages = "085110",
year = "2015",
title = "Large-deviation joint statistics of the finite-time {Lyapunov} spectrum in isotropic turbulence",
author = "P. L. Johnson and C. Meneveau",
doi = {https://doi.org/10.1063/1.4928699}
}

@article{zrms84,
journal = "J. Fluid Mech.",
volume = "144",
pages = "1",
year = "1984",
author = "Ya. B. Zel'dovich and A. A. Ruzmaikin and S. A. Molchanov and D. D. Sokoloff",
title = "Kinematic dynamo problem in a linear velocity field",
doi = {https://doi.org/10.1017/S0022112084001488}
}

@article{fma07,
author = "G. Falkovich and M. {Martins Afonso}",
journal = "Phys. Rev. {\rm E}",
volume = "76",
pages = "026312 ",
year = "2007",
title = "Fluid-particle separation in a random flow described by the telegraph model",
doi = {https://doi.org/10.1103/PhysRevE.76.026312}
    }

@article{cfvp91,
journal = "Riv. Nuovo Cimento",
volume = "14",
pages = "1",
year = "1991",
author = "A. Crisanti and M. Falcioni and A. Vulpiani and G. Paladin",
title = "Lagrangian chaos: Transport, mixing and diffusion in fluids",
doi = {https://doi.org/10.1007/BF02811193}
}

@article{ckv98,
journal = "Phys. Rev. Lett.",
volume = "80",
pages = "512",
year = "1998",
author = "M.~Chertkov and I.~Kolokolov and M.~Vergassola",
title = "Inverse versus Direct Cascades in Turbulent Advection",
doi = {https://doi.org/10.1103/PhysRevLett.80.512}
}

@article{gm13,
author = "K. Gustavsson and B. Mehlig",
title = "Lyapunov Exponents for Particles Advected in Compressible Random Velocity Fields at Small and Large {K}ubo Numbers",
journal = "J. Stat. Phys.",
volume = "153",
pages = "813--827",
year = "2013",
doi = {https://doi.org/10.1007/s10955-013-0848-z}
    }

@article{v21,
author = "D. Vincenzi",
title = "Effect of internal friction on the coil--stretch transition in turbulent flows",
journal = "Soft Matter",
volume = "17",
pages = "2421--2428",
year = "2021",
doi = {https://doi.org/10.1039/D0SM01981J}
    }

@article{dmv14,
journal = "J. Fluid Mech.",
volume = "761",
pages = "431",
year = "2014",
author = "A.~Dhanagare and S.~Musacchio and D.~Vincenzi",
title = "Weak--strong clustering transition in renewing compressible flows",
doi = {https://doi.org/10.1017/jfm.2014.634}
}

@article{bppv85,
  title = {Characterisation of intermittency in chaotic systems},
  author = {R.~Benzi and G.~Paladin and G.~Parisi and A.~Vulpiani},
  journal = {J. Phys. {\rm A:} Math.~Gen.},
  volume = {18},
  pages = {2157},
  year = {1985},
    url = {https://dx.doi.org/10.1088/0305-4470/18/16/028}
}

@article{mv11,
journal = "J. Fluid Mech.",
volume = "670",
pages = "326--336",
year = "2011",
author = "S.~Musacchio and D.~Vincenzi",
title = "Deformation of a flexible polymer in a random flow with long correlation time",
doi ={https://doi.org/10.1017/S0022112010006385}
}

@article{bbbcmt06,
title = "{Lyapunov} exponents of heavy particles in turbulence",
journal = "Phys. Fluids",
volume = "18",
pages = "091702",
year = "2006",
author = "J. Bec and L. Biferale and G. Boffetta and M. Cencini and S. Musacchio and F. Toschi",
doi = {https://doi.org/10.1063/1.2349587}
}

@article{orl82,
journal = "J. Non-Newtonian Fluid Mech.",
volume = "10",
pages = "291--318",
year = "1982",
author = "W.~L.~Olbricht and J.~M.~Rallison and L.~G.~Leal",
title="Strong flow criteria based on microstructure deformation",
doi = {https://doi.org/10.1016/0377-0257(82)80006-2}
}

@article{coletti_jfm, title={Behaviour of finite-size floating particles in free-surface turbulence}, volume={1019}, DOI={10.1017/jfm.2025.10648}, journal={J. Fluid Mech.}, author={Sanness Salmon, H. R. and Chang, K. and Mucignat, C. and Coletti, F.}, year={2025}, pages={A61}}

@article{fgv01,
journal = "Rev. Mod. Phys.",
volume = "73",
pages = "913--975",
year = "2001",
author = "G.~Falkovich and K.~Gaw\c{e}dki and M.~Vergassola",
title = "Particles and fields in fluid turbulence",
doi = {https://doi.org/10.1103/RevModPhys.73.913}
}

@article{c00,
journal = "Phys. Rev. Lett.",
volume = "84",
pages = "4761--4764",
year = "2000",
title = "Polymer stretching by turbulence",
author = "M.~Chertkov",
doi = {https://doi.org/10.1103/PhysRevLett.84.4761}
}

@book{bird,
  author    = {R.~B.~Bird and C.~F.~Curtiss and R.~C.~Armstrong and O.~Hassager},
  title     = {Dynamics of Polymeric Liquids},
  publisher = {Wiley},
  year      = 1987,
  volume    = 2,
}

@book{bjpv98,
  author    = {T. Bohr and M. H. Jensen and G. Paladin and A. Vulpiani},
  title     = {Dynamical Systems Approach to Turbulence},
  publisher = {Cambridge University Press},
  year      = 1998,
address = {Cambridge, UK},
doi = {https://doi.org/10.1017/CBO9780511599972}
}

@book{ccv2010,
  author    = {F.~Cecconi and M.~Cencini and A.~Vulpiani}, 
  title     = {Chaos: from Simple Models to Complex Systems},
  publisher = {World Scientific},
  year      = 2010,
  address   = {Singapore},
}

@book{o96,
  author    = {H. C. \"Ottinger}, 
  title     = {Stochastic Processes in Polymeric Fluids},
  publisher = {Springer},
  year      = 1996,
  address   = {Berlin},
doi = {https://doi.org/10.1007/978-3-642-58290-5}
}

@article{bcm03,
journal = "Phys. Rev. Lett.",
volume = "91",
pages = "034501",
year = "2003",
title = "Two-dimensional turbulence of dilute polymer solutions",
author = "G. Boffetta and A. Celani and S. Musacchio",
doi = {https://doi.org/10.1103/PhysRevLett.91.034501}
}

@article{bmpb12,
journal = "Phys. Rev. {\rm E}",
volume = "86",
pages = "056314",
year = "2012",
title = "Statistics of polymer extensions in turbulent channel flow",
author = "F. Bagheri and D. Mitra and P. Perlekar and L. Brandt",
doi = {https://doi.org/10.1103/PhysRevE.86.056314}
}

@article{gcs05,
journal = "Europhys. Lett.",
volume = "71",
pages = "221--227",
year = "2005",
title = "Single-polymer dynamics: Coil-stretch transition in a random flow",
author = "S Gerashchenko and C. Chevallard and V. Steinberg",
url = {https://dx.doi.org/10.1209/epl/i2005-10087-1}
}

@article{bfl00,
journal = "Phys. Rev. Lett.",
volume = "84",
pages = "4765",
year = "2000",
title = "Turbulent dynamics of polymer solutions",
author = "E. Balkovsky and A. Fouxon and V. Lebedev",
doi = {https://doi.org/10.1103/PhysRevLett.84.4765}
}

@article{fak19,
doi = {10.1088/1742-5468/ab3458},
url = {https://dx.doi.org/10.1088/1742-5468/ab3458},
year = {2019},
month = {aug},
publisher = {IOP Publishing and SISSA},
volume = {2019},
number = {8},
pages = {083211},
author = {Fouxon, I. and Ainsaar, S. and Kalda, J.},
title = {Quartic polynomial approximation for fluctuations of separation of trajectories in chaos and correlation dimension},
journal = {J. Stat. Mech.: Theory Exp.}
}

@article{fouxon2023,
journal = "Phys. Fluids",
volume = "35",
pages = "125114",
year = "2023",
author = "H. Yu and I. Fouxon and J. Wang and X. Li and L. Yuan and S. Mao and M. Mond ",
title = "{Lyapunov exponents and Lagrangian chaos suppression in compressible homogeneous isotropic turbulence}",
doi = {https://doi.org/10.1063/5.0175016}
}

@article{bgh04,
title = {Multifractal Clustering in Compressible Flows},
  author = {Bec, J. and Gaw\ifmmode \mbox{\c{e}}\else \c{e}\fi{}dzki, K. and Horvai, P.},
  journal = {Phys. Rev. Lett.},
  volume = {92},
  issue = {22},
  pages = {224501},
  numpages = {4},
  year = {2004},
  month = {Jun},
  publisher = {American Physical Society},
  doi = {10.1103/PhysRevLett.92.224501},
  url = {https://link.aps.org/doi/10.1103/PhysRevLett.92.224501}
}

@incollection{b91,
	author = "P. Baxendale",
	title = "Statistical Equilibrium and Two-Point Motion for a Stochastic Flow of Diffeomorphisms",
	booktitle = "Spatial Stochastic Processes",
	year = "1991",
	editor = "K. S. Alexander and J. C. Watkins",
	pages = "189--218",
	publisher = "Birkhäuser",
	series = "Progress in Probability",
	volume = "19",
	address = "Boston",
doi = {https://doi.org/10.1007/978-1-4612-0451-0}
}

@article{bfl01,
journal = "Phys. Rev. {\rm E}",
volume = "64",
pages = "056301",
year = "2001",
author = "E. Balkovsky and A. Fouxon and V. Lebedev",
title = "Turbulence of polymer solutions",
doi = {https://doi.org/10.1103/PhysRevE.64.056301}
}

@article{wg10,
journal = "Phys. Rev. {\rm E}",
volume = "81",
pages = "066301",
year = "2010",
title = "Coil-stretch transition in an ensemble of polymers in isotropic turbulence",
author = "T. Watanabe and T. Gotoh",
doi  = {https://doi.org/10.1103/PhysRevE.81.066301}
}

@article{pav16,
    journal = "Phys. Rev. {\rm E}",
    volume = "94",
    pages = "020501(R)",
    year = "2016",
    author = "E. L. C. VI M. Plan and A. Ali and D. Vincenzi",
    title = "Bead-rod-spring models in random flows",
    doi    = {https://doi.org/10.1103/PhysRevE.94.020501}
}

@article{vwrp21,
journal = "J. Fluid Mech.",
volume = "912",
pages = "A18",
year = "2021",
author = "D. Vincenzi and T. Watanabe and S. S. Ray and J. Picardo",
Title = "Polymer scission in turbulent flows",
doi = {https://doi.org/10.1017/jfm.2020.1092}, 
}

@article{ravichandran-rev2017,
title = {Clustering of heavy particles in vortical flows: a selective review},
journal = {S\={a}dhan\={a}},
author = {S. Ravichandran and P. Deepu and R. Govindarajan},
volume = {42},
number = {4},
pages = {597-605},
year = {2017},
doi = {https://www.ias.ac.in/article/fulltext/sadh/042/04/0597-0605}
}

@article{Haller08,
title = {Where do inertial particles go in fluid flows?},
journal = {Physica {\rm D}},
volume = {237},
number = {5},
pages = {573-583},
year = {2008},
author = {G. Haller and T. Sapsis},
doi = {https://doi.org/10.1016/j.physd.2007.09.027}
}

@article{Boffetta07-eulerian,
year = {2007},
month = {mar},
publisher = {},
volume = {78},
number = {1},
pages = {14001},
author = {Boffetta, G. and Celani, A. and {De Lillo}, F. and Musacchio, S.},
title = {The {Eulerian} description of dilute collisionless suspension},
journal = {Europhys. Lett.},
doi = {10.1209/0295-5075/78/14001}
}

@article{bec21-dusty,
    author = {J. Bec and F. Laenen and S. Musacchio},
    title = {Dusty turbulence},
    year = {2017},
    journal = {\href{https://doi.org/10.48550/arXiv.1702.06773}{arXiv:1702.06773}}
}

@article{Celani2013-colored-noise,
  title = {White-noise limit of nonwhite nonequilibrium processes},
  author = {Bo, S. and Celani, A.},
  journal = {Phys. Rev. {\rm E}},
  volume = {88},
  issue = {6},
  pages = {062150},
  numpages = {7},
  year = {2013},
  month = {Dec},
  publisher = {American Physical Society},
  doi = {10.1103/PhysRevE.88.062150},
  url = {https://link.aps.org/doi/10.1103/PhysRevE.88.062150}
}

\end{document}